\documentclass[12pt]{article}

\usepackage{epsfig}
\usepackage{amssymb,amsmath}
\usepackage{multirow}
\usepackage{rotating}
\usepackage{graphicx}
\usepackage{lscape}
\usepackage{hyperref}
\usepackage{xcolor}
\usepackage{url}

\usepackage{comment}

\usepackage{cite} % to cite multiple reference in the same brackets in short way.

\newcommand{\bea}{\begin{eqnarray}}
\newcommand{\eea}{\end{eqnarray}}

 \makeatletter
\def\alt{\mathrel{\mathpalette\gl@align<}}
\def\agt{\mathrel{\mathpalette\gl@align>}}
\def\gl@align#1#2{\lower.6ex\vbox{\baselineskip\z@skip\lineskip\z@
\ialign{$\m@th#1\hfil##\hfil$\crcr#2\crcr\sim\crcr}}} \makeatother

\begin{document}

\begin{flushright}
%preprint number \\
\end{flushright}

\vspace*{1.0cm}

\begin{center}
\baselineskip 20pt 
{\Large\bf 
%$b \rightarrow s \gamma$ transition in
%a model with vectorlike fermions \\
%and secluded scalars \\
%: (present and prospect)
%
%B\'elanger$-$Delaunay$-$Westhoff model
%
%$b \rightarrow s$ transitions and CKM unitarity %violation 
%from vectorlike fermions
%with $U(1)_X$ charges \\
%prospect at Belle II \\

A model with vectorlike fermions and $U(1)_X$ symmetry: CKM unitarity, $b \rightarrow s$ transitions,\\
and prospect at Belle II
}
\vspace{1cm}

{\large 
Sang Quang Dinh$^{a}$
and 
Hieu Minh Tran$^{b,}$\footnote{ E-mail: hieu.tranminh@hust.edu.vn} 
} 
\vspace{.5cm}

{\baselineskip 20pt \it
$^a$VNU University of Science, Vietnam National University - Hanoi,  \\
334 Nguyen Trai Road, Hanoi, Vietnam \\
\vspace{2mm} 
$^b$Hanoi University of Science and Technology, 1 Dai Co Viet Road, Hanoi, Vietnam 
}

\vspace{.5cm}

\vspace{1.5cm} {\bf Abstract}
\end{center}

The updated analysis of the LHCb Collaboration on the lepton flavor violation suggests that the new physics should couple to muons and electrons with comparable magnitudes, resulting in the anomalies in both rare decay channels,
$b \rightarrow s \mu^+ \mu^-$ and
$b \rightarrow s e^+ e^-$.
Meanwhile, the recent result of the Muon $g-2$ experiment with higher precision has increased the existing tension with the standard model prediction.
In this paper, we consider an extension of the standard model with a new sector consisting of vectorlike fermions and two scalar charged under an extra $U(1)_X$ gauge symmetry.
The exotic Yukawa interactions in the this model lead to the quark mixing responsible for the additional contributions to the flavor changing neutral currents in $B$-meson decays, 
and solve the muon $g-2$ discrepancy.
We derive the analytic expression of the new physics contributions to the Wilson coefficient $C_7$ in the effective Hamiltonian, 
and point out that the CKM unitarity violation can be explained within this context.
By calculating the branching ratio of the inclusive radiative $B$ decay, the impact of current experimental data of the $b \rightarrow s \gamma$ transition on the model and the future prospect at the Belle II experiment are investigated.
Taking into account the current data on 
the muon anomalous magnetic moment,
the CKM unitarity violation, the constraints on the flavor observables relevant to the $b \rightarrow s$ transitions, 
the LHC searches for vectorlike quarks,
and the perturbation limits of the couplings,
the viable parameter regions of the model are identified.

%To address the muon $g-2$ deviation and the 
%anomalies of
% the semileptonic decays of $B$ mesons, 
%B\'elanger, Delaunay, and Westhoff introduced a new sector consisting of vectorlike fermions and two scalar charged under an extra $U(1)_X$ gauge symmetry.
%%
%The exotic Yukawa interactions in this model lead to the quark mixing responsible for the additional contributions to the flavor changing neutral currents in $B$-meson decays.
%%
%In this paper, we derive the analytic expression of the new physics contributions to the Wilson coefficient $C_7$ in the effective Hamiltonian.
%%
%By calculating the branching ratio of the inclusive radiative $B$ decay, the impact of current experimental data of the $b \rightarrow s \gamma$ transition on the model and the future prospect at the Belle II experiment are investigated.
%%
%Taking into account the recent data on 
%the muon anomalous magnetic moment,
%the CKM unitarity violation, the updated constraints on the flavor observables relevant to the $b \rightarrow s$ transitions, and the perturbation limits of the couplings,
%the viable parameter regions of the model are identified.

%============================
%
%The constraints on the semileptonic decays of $B$ mesons are also taken into account.
%
%
%LEP and LHC constraints in the searches for $Z'$ boson

\thispagestyle{empty}

%\bigskip
\newpage

\addtocounter{page}{-1}

%%%%%%%%%%%%%%%%%%%%%%%%%%
%\baselineskip 36pt
% Main body
%%%%%%%%%%%%%%%%%%%%%%%%%%
\baselineskip 18pt
%%%%%%%%%%%%%%%%%%%%%%%%%%

%\begin{fmffile}{feyndiagrams}
%to generate unique file "feyndiagrams.mp" to run with "mpost" command (avoiding multiple .mp files)

%%%%%%%%%%%%%%%%%%%%%%%%%%
\section{Introduction}  
%%%%%%%%%%%%%%%%%%%%%%%%%%

Although the standard model (SM) is in agreement with most of the experimental results, there are numerous evidences showing that it is not enough to explain them, ranging from cosmological observations to measurements at colliders.
Therefore, this model is considered as an effective theory, and the new physics is likely somewhere around the corner.
Recently, the updated determination of the Cabibbo-Kobayashi-Maskaw (CKM) matrix elements showed that there is a 2.2$\sigma$ deviation from unitary in the first row of this mixing matrix
\cite{Workman:2022ynf}:
\begin{eqnarray}
|V_{ud}|^2 + |V_{us}|^2 + |V_{ub}|^2 &=& 
	0.9985 \pm 0.0007.
\label{CKM_unitarity}
\end{eqnarray}
This CKM anomaly implies a non-negligible effect of new physics   \cite{Belfatto:2019swo}.
Beside the violation of the CKM unitarity, 
other constraints such as those on the 
muon $g-2$ 
and the anomalies in the semileptonic B decays
also indicate the existence of new physics
\cite{Arnan:2019uhr, Crivellin:2015mga}.
After the 2023 result of the Muon $g-2$ experiment at Fermilab, the deviation of the muon anomalous magnetic moment between the world average value and the SM prediction%
\footnote{Here, we take into account the result in the Ref. \cite{Whitepaper2020}, although there are other approaches
\cite{Borsanyi:2020mff, CMD-3:2023alj} showing different SM predictions of the muon $g-2$ with less tension.
}
is increased to be more than 5$\sigma$ \cite{Muong-2:2023cdq}, 
mostly due to the reduction of the experimental uncertainties.
Moreover, the discrepancies between the SM predictions and the experimental values for the branching ratios of the processes
$B^+ \rightarrow K^+ \mu^+ \mu^-$, 
$B^0 \rightarrow K^0 \mu^+ \mu^-$, and
$B^+ \rightarrow K^+ e^+ e^-$
are reported to be 4.2$\sigma$, 3.1$\sigma$, and 2.7$\sigma$ respectively
\cite{Parrott:2022zte}. 
On the other hand, the updated analysis of the LHCb Collaboration \cite{LHCb:2022qnv} showed the good agreement between the measured lepton universality observables, 
$R_K$ and $R_{K^*}$, and their SM predictions.
This implies that the new physics may couple to muons and electrons with comparable magnitudes 
\cite{Alguero:2023jeh, Wen:2023pfq, Allanach:2023uxz}, 
suggesting that the deviation for 
BR($B^+ \rightarrow K^+ e^+ e^-$) is even more severe than the above value \cite{pkoppenb}.

Among the sensitive probes to the new physics beyond the SM, the 
$b \rightarrow s \gamma$ transition is of particular importance.
Being a flavor changing neutral current (FCNC) process, this transition is forbidden at the tree level and only arises due to quantum corrections at loop levels in the SM.
Interestingly, its rate is of order $G_F^2 \alpha$ that is larger than the rates of most of other FCNC processes 
\cite{Buras:1993xp, Buchalla:1995vs}.
At the hadronic level, this transition is identified as the $B \rightarrow X_s \gamma$ decay, of which the branching ratio for $E_\gamma > 1.6$ GeV
 predicted by the SM is given by 
\cite{Chetyrkin:1996vx, Misiak:2006zs, Misiak:2015xwa, Czakon:2015exa, Misiak:2020vlo}
\begin{eqnarray}
BR(B \rightarrow X_s \gamma)_\text{SM}	&=& 
	(3.40 \pm 0.17) \times 10^{-4}.
\end{eqnarray}
The measurements of this inclusive decay process have been performed by several experiments including
CLEO \cite{CLEO:2001gsa}, 
BaBar \cite{BaBar:2007yhb, BaBar:2012eja, BaBar:2012fqh}, and
Belle \cite{Belle:2009nth, Belle:2014nmp}.
The world average value of the branching ratio was evaluated by the Heavy Flavor Averaging Group \cite{HFLAV:2022pwe} as
\begin{eqnarray}
BR(B \rightarrow X_s \gamma)_\text{exp}	&=& 
	(3.49 \pm 0.19) \times 10^{-4},
\end{eqnarray}
for the same cut on the radiated photon energy.
The good agreement between the theoretical prediction and the measurement implies that the contributions of new physics to the $b \rightarrow s \gamma$ decay should not be too large.
In the near future, when the relative uncertainty is reduced to the level of a few per cent with the luminosity of 50 ab$^{-1}$ at the Belle II experiment
\cite{Belle-II:2022cgf, DiCanto:2022icc}, it is expected that the upcoming data will strongly constrain the new physics contribution to this process if the center value of the decay rate remains unchanged.

It has been shown that many new physics models are strictly constrained by the $b \rightarrow s \gamma$ transition,
see for example Refs. 
\cite{Chang:2000gz, Akeroyd:2001gf,
Idarraga:2005ia, Lunghi:2007ak, Branco:2011iw, Hermann:2012fc, Jung:2012vu, Crivellin:2013wna, Das:2015kea, Misiak:2017bgg, Haller:2018nnx, Arco:2020ucn, Atkinson:2021eox, Arco:2022xum, Enomoto:2022rrl,
Akeroyd:2020nfj,
Bertolini:1990if, Barbieri:1993av, Borzumati:1994te, Degrassi:2000qf, Carena:2000uj, Demir:2001yz, Baek:2002wm, Hurth:2003vb, Ellis:2006ix, Gomez:2006uv, Ellis:2007fu, Heinemeyer:2008fb, Olive:2008vv, Okada:2010xe, Zhang:2014nya, GAMBIT:2017snp, Yang:2018fvw,
Haisch:2007vb, Freitas:2008vh, Moch:2015oka, Blanke:2012tv, Datta:2016flx,
Cheung:2017efc,
Aliev:1996cyj,
NguyenTuan:2020xls,
Gabrielli:2016cut,
Aoki:2000ze, Aoki:2001xr, Morozumi:2018cnc, Vatsyayan:2020jan}.
Among those, models with vectorlike quarks turn out to be interesting since they can naturally address the CKM unitarity violation by the mixing between the SM and the vectorlike quarks  \cite{Kawamura:2019rth, Cheung:2020vqm, Crivellin:2020ebi, Cherchiglia:2021vhe, Belfatto:2021jhf, Branco:2021vhs, Balaji:2021lpr, CarcamoHernandez:2021yev, Accomando:2022ouo, Guedes:2022cfy, Branco:2022fmj}.
%
%
%
%
%- Model with vector quarks: \cite{Chang:2000gz, Akeroyd:2001gf}
%
%* THDM: \cite{Hewett:1992is,Idarraga:2005ia, Lunghi:2007ak, Branco:2011iw, Hermann:2012fc, Jung:2012vu, Crivellin:2013wna, Das:2015kea, Misiak:2017bgg, Haller:2018nnx, Arco:2020ucn, Atkinson:2021eox, Arco:2022xum, Enomoto:2022rrl}
%
%* 3HDM: \cite{Akeroyd:2020nfj}
%
%* SUSY: \cite{Bertolini:1990if, Barbieri:1993av, Borzumati:1994te, Degrassi:2000qf, Carena:2000uj, Demir:2001yz, Baek:2002wm, Hurth:2003vb, Ellis:2006ix, Gomez:2006uv, Ellis:2007fu, Heinemeyer:2008fb, Olive:2008vv, Okada:2010xe,
%Zhang:2014nya, GAMBIT:2017snp, Yang:2018fvw}
%
%* Extra dimensions: 
%\cite{Haisch:2007vb, Freitas:2008vh, Moch:2015oka, Blanke:2012tv, Datta:2016flx}
%
%* Leptoquarks: \cite{Cheung:2017efc}
%
%* Four generation: \cite{Aliev:1996cyj}
%
%* 3-3-1-1 model: \cite{NguyenTuan:2020xls}
%
%* Vectorlike fermion: \cite{Aoki:2000ze, Aoki:2001xr, Morozumi:2018cnc, Vatsyayan:2020jan}
%
%
%
%
In this paper, we are interested in the model with additional vectorlike fermions and a secluded scalar sector charged under an extra Abelian gauge symmetry $U(1)_X$ proposed 
by B\'elanger, Delaunay, and Westhoff
in Refs.
\cite{Belanger:2015nma, Belanger:2016ywb}.
Due to the exotic Yukawa coupling between the muon, the vectorlike lepton and the scalar, of which the field value vanishes at the minimum of the scalar potential, the model can explain the measured muon anomalous magnetic moment.
The other type of Yukawa couplings between the SM quarks, the vectorlike ones and another scalar that develops nonzero vacuum expectation value leads to the mixing in the quark sector among the SM and the vectorlike quarks.
This is the source for the the additional contributions to the FCNC processes such as the semileptonic decays of $B$ mesons.
The new physics contribution Wilson coefficients $C^{(')}_{9,10}$ were calculated analytically at the leading order in Ref. \cite{Dinh:2020inx}
in the general case with a non-vanishing gauge kinetic mixing term, allowing the evaluation of multiple relevant flavor observables such as the decay rates of the $b \rightarrow s \ell^+ \ell^-$ processes.
In this follow-up paper, we investigate the ability of this model to reconcile both the constraints on the CKM unitarity violation and the $b \rightarrow s \gamma$ decay, 
while keeping other predicted observables consistent with their experimental values.
Taking into consideration the recent LHC searches for vectorlike quarks,
the future prospect of this model at the Belle II experiment will be discussed as well.

The structure of the paper is as follows.
In Section 2, the setup of the model is briefly reviewed.
In Section 3, the analytic expression of the new physics contributions to the Wilson coefficient $C_7$ in the effective Hamiltonian is derived.
In Section 4, the numerical analyses are carried out.
The dependence of the branching ratio of $b\rightarrow s \gamma$ decay on the input parameters is presented. 
Taking into account various constraints, we then identify the viable parameter regions of the model.
Here, the impact of the expected result at the Belle II experiment is also considered.
The last section is devoted to the conclusion.

%=========================================

%\begin{eqnarray}
%\text{BR}(b\rightarrow s \gamma)_\text{Belle2} &=& (3.490 \pm 0.698) \times 10^{-4}.
%\end{eqnarray}

%Brief review of recent progress on $b \rightarrow s \gamma$

%%%%%%%%%%%%%%%%%%%%%%%%%%%%%%
\section{The model}
%%%%%%%%%%%%%%%%%%%%%%%%%%%%%%

In the considered model, beside the SM particles, the new particles introduced  are the vectorlike lepton and quark doublets of the gauge group $SU(2)_L$,
\begin{eqnarray}
L_{L,R} = 
	\begin{pmatrix}
	N_{L,R} \\ E_{L,R}
	\end{pmatrix}	,	\qquad
Q_{L,R} = 
	\begin{pmatrix}
	U_{L,R} \\ D_{L,R}
	\end{pmatrix}	,
\end{eqnarray}
and two complex scalars, $\chi$ and $\phi$, that are singlets under the SM gauge groups.
The symmetry of this model is 
$SU(3)_C \otimes SU(2)_L \otimes U(1)_Y \otimes U(1)_X$
that extends the SM symmetry by adding an extra Abelian gauge group.
The SM particles are invariant under $U(1)_X$ transformation, while the new particles transform nontrivially with the $U(1)_X$ charges given in Table \ref{NP} together with other properties.

%%%%%%%%%%%%%%%%%%%%%

\begin{table}[h]
\caption{Properties of new particles introduced in the model \cite{Belanger:2015nma}.}
\label{NP}
\begin{center}
\begin{tabular}{|c|c|c|c|c|c|}
\hline
Particles	&	Spin	&	$SU(3)_C$	&	$SU(2)_L$	&	$U(1)_Y$	&	$U(1)_X$	\\
\hline
$L_L, L_R$	&	1/2	&	\textbf{1}	&	\textbf{2}	&	-1/2	&	1	\\
$Q_L, Q_R$	&	1/2	&	\textbf{3}	&	\textbf{2}	&	1/6	&	-2	\\
\hline
$\chi$		&	0	&	\textbf{1}	&	\textbf{1}	&	0	&	-1	\\
$\phi$		&	0	&	\textbf{1}	&	\textbf{1}	&	0	&	2	\\
\hline
\end{tabular}
\end{center}
\end{table}
%%%%%%%%%%%%%%%%%%%%%

The Lagrangian consists of the SM part and the part involving new physics:
\begin{eqnarray}
\mathcal{L} &=& \mathcal{L}_\text{SM} + \mathcal{L}_\text{NP},
\end{eqnarray}
where
\begin{eqnarray}
\mathcal{L}_\text{NP} & \supset &
	- \; \lambda_{\phi H} |\phi|^2 |H|^2
	- \lambda_{\chi H} |\chi|^2 |H|^2
	- \left[
	y \overline{\ell_L} L_R \chi +
	w \overline{q_L} Q_R \phi + h.c.
	\right] 
	- V_0(\phi,\chi) \nonumber	\\
&&
	- \; ( M_L \overline{L_L} L_R
	+ M_Q \overline{Q_L} Q_R	+ h.c. ) \, .
\label{Lnp}
\end{eqnarray}
In this equation, the SM left-handed lepton and quark doublets are denoted as
\begin{eqnarray}
\ell_L^i = 
	\begin{pmatrix}
	\nu^e_L \\ e_L
	\end{pmatrix}_i ,	\qquad
q_L^i =
	\begin{pmatrix}
	u_L	\\ d_L
	\end{pmatrix}_i ,	\qquad
(i = 1,2,3) .
\end{eqnarray}
The mass terms of the vectorlike leptons and quarks are allowed in the Lagrangian (\ref{Lnp}) with the corresponding mass matrices:
\begin{eqnarray}
M_L = 
	\begin{pmatrix}
	m_N	&	0	\\
	0	&	m_E	
	\end{pmatrix} , 
	\quad
M_Q	=
	\begin{pmatrix}
	m_U	&	0	\\
	0	&	m_D	
	\end{pmatrix}	.
\end{eqnarray}
In our analysis, for simplicity, we assume the mass degeneration in the two vectorlike doublets, i.e.
$m_N = m_E = m_L$,
$m_U = m_D = m_Q$.
The scalar potential $V_0(\chi,\phi)$ is given explicitly by
\begin{eqnarray}
V_0(\chi,\phi) &=&
	\lambda_\phi |\phi|^4 + m^2_\phi |\phi|^2 +
	\lambda_\chi |\chi|^4 + m^2_\chi |\chi|^2 + 
	\lambda_{\phi\chi} |\phi|^2 |\chi|^2 +
	\left( r \phi \chi^2 + h.c. \right) .
\label{V0}
\end{eqnarray}

We assume that the gauge group $U(1)_X$ is spontaneously broken by the vacuum expectation value (VEV) of the scalar $\phi$,
\begin{eqnarray}
\langle \phi \rangle &=& 		
	\sqrt{\frac{-m'^2_\phi}	
		{2\lambda_\phi}}	,
\end{eqnarray}
while the other scalar $\chi$ does not develop a nonzero VEV.
In the above equation, we denote
\begin{eqnarray}
m'^2_\phi &=& m^2_\phi + \lambda_{\phi H} \langle H \rangle^2	,
\end{eqnarray}
where $\langle H \rangle = 174$ GeV is the VEV of the SM Higgs field.
Due to the nonzero VEV, $\langle \phi \rangle$, the $U(1)_X$ gauge boson $Z'$ acquires a mass
\begin{eqnarray}
m_{Z'} &=& 2\sqrt{2} g_X \langle \phi \rangle	.
\label{mZp}
\end{eqnarray}
with $g_X$ being the $U(1)_X$ gauge coupling.

Decomposing the complex scalar field $\phi$ into the real and imaginary components,
\begin{eqnarray}
\phi &=& 
	\langle \phi \rangle + 
	\frac{1}{\sqrt{2}}
	\left( \varphi_r + i \varphi_i \right)	,
\label{phi}
\end{eqnarray} 
their masses are respectively found to be
\begin{eqnarray}
m_{\varphi_r} &=& 
	2 \sqrt{\lambda_\phi} \langle \phi \rangle,	\\
m_{\varphi_i} &=& 	0	\,	.
\end{eqnarray}
In the unitary gauge, the massless Nambu-Goldstone boson $\varphi_i$ is absorbed by the $Z'$ gauge boson.
For the scalar field $\chi$, by the similar decomposition
\begin{eqnarray}
\chi &=& 
	\frac{1}{\sqrt{2}}
	\left( \chi_r + i \chi_i \right)	\, ,
\end{eqnarray} 
the $2\times 2$ mass matrix for these real component fields is derived as
\begin{eqnarray}
\frac{1}{2}
	\begin{pmatrix}
	\chi_r	&	\chi_i
	\end{pmatrix}
M^2_\chi
	\begin{pmatrix}
	\chi_r	\\	\chi_i
	\end{pmatrix}
&=&
\frac{1}{2}
	\begin{pmatrix}
	\chi_r	&	\chi_i
	\end{pmatrix}
	\begin{pmatrix}
	m'^2_\chi + (r+r^*) \langle \phi \rangle	&	i(r-r^*) \langle \phi \rangle	\\
	i(r-r^*) \langle \phi \rangle	&	m'^2_\chi - (r+r^*) \langle \phi \rangle
	\end{pmatrix}
	\begin{pmatrix}
	\chi_r	\\	\chi_i
	\end{pmatrix}	,
\end{eqnarray}
where
\begin{eqnarray}
m'^2_\chi	&=&
	m^2_\chi + 
	\lambda_{\chi H} \langle H \rangle^2 +
	\lambda_{\phi\chi} \langle \phi \rangle^2	.
\end{eqnarray}
In the case that the coupling $r$ is real, the matrix $M_\chi^2$ is diagonal, and the masses of the particles $\chi_r$ and $\chi_i$ are respectively
\begin{eqnarray}
m_{\chi_r}^2	&=&	m'^2_\chi + 2r \langle \phi \rangle	,	\\
m_{\chi_i}^2	&=&	m'^2_\chi - 2r \langle \phi \rangle	.
\end{eqnarray}

%Mixing between vector-like and SM fermions

In the lepton sector, there is no mass mixing between the SM leptons and the vectorlike ones because 
$\langle \chi \rangle =0$.
However, in the quark sector, the VEV of $\phi$ generates mass mixing terms among the SM quarks and the vectorlike ones via the new exotic Yukawa interactions with the couplings 
$w = ( w_1, w_2, w_3 )$ in Eq. (\ref{Lnp}).
To diagonalize the quark mass matrices, $M^u$ and $M^d$, four $4\times 4$ unitary matrices are necessary to transform the quark gauge eigenstates, 
$(u^1, u^2, u^3, U)$ and 
$(d^1, d^2, d^3, D)$,
into the mass eigenstates, 
$(u, c, t, \mathcal{U})$ and 
$(d, s, b, \mathcal{D})$:
\begin{eqnarray}
\begin{pmatrix}
u_{L,R} \\	c_{L,R} \\	t_{L,R} \\	\mathcal{U}_{L,R}
\end{pmatrix}	=
	\left( V^u_{L,R} \right)_{4 \times 4}
	\begin{pmatrix}
	u_{L,R}^1 \\	u_{L,R}^2 \\
	u_{L,R}^3 \\	U_{L,R}	
	\end{pmatrix}	,	\qquad
\begin{pmatrix}
d_{L,R} \\	s_{L,R} \\	b_{L,R} \\	\mathcal{D}_{L,R}
\end{pmatrix}	=
	\left( V^d_{L,R} \right)_{4 \times 4}
	\begin{pmatrix}
	d_{L,R}^1 \\	d_{L,R}^2 \\
	d_{L,R}^3 \\	D_{L,R}	
	\end{pmatrix}	.
\label{qmix}
\end{eqnarray}
As a consequence, the diagonal mass matrices of the up-type and down-type quarks are then given by
\begin{eqnarray}
M^u_\text{diag}	&=&	V^u_L M^u (V^u_R)^\dagger	,	\\
M^d_\text{diag}	&=&	V^d_L M^d (V^d_R)^\dagger	.
\end{eqnarray}

%%%%%%%%%%%%%%%%%%%%%%%%%%%%%%%
\section{New physics contributions to the Wilson coefficient $C_7$}
%%%%%%%%%%%%%%%%%%%%%%%%%%%%%%%

The effective Hamiltonian describing the $ b \rightarrow s \gamma$ transitions is given by
\cite{Buchalla:1995vs, Greub:1996tg}
\begin{eqnarray}
\mathcal{H}_{\text{eff}}&=&
	-\frac{4 G_F}{\sqrt{2}}V_{tb}V^*_{ts}\sum_{i=1}^{8} (C_i \mathcal{O}_i + C_i' \mathcal{O}_i') + h.c.
\end{eqnarray}
Here, the operators most directly relevant to the $b\rightarrow s \gamma$ process are:
\begin{eqnarray}
\mathcal{O}_7 &=&
	\frac{e}{16\pi^{2}} m_b \left( \overline{s_L} \sigma^{\mu\nu} b_R \right) F_{\mu\nu} , \\
\mathcal{O}'_7 &=&
	\frac{e}{16\pi^{2}} m_s \left( \overline{s_R} \sigma^{\mu\nu} b_L \right) F_{\mu\nu} .
\end{eqnarray}
Due to the suppression factor $\frac{m_s}{m_b} \ll 1$ emerging from the mass-insertion on the external $s$ quark, the contribution of the operator $\mathcal{O}'_7$ is negligible in comparison to that of $\mathcal{O}_7$.

%%%%%%%%%%%%%%%%%%%%%%%%%%%%%%%%%%
\begin{figure}[h!]
\begin{center}
\includegraphics[scale=1]{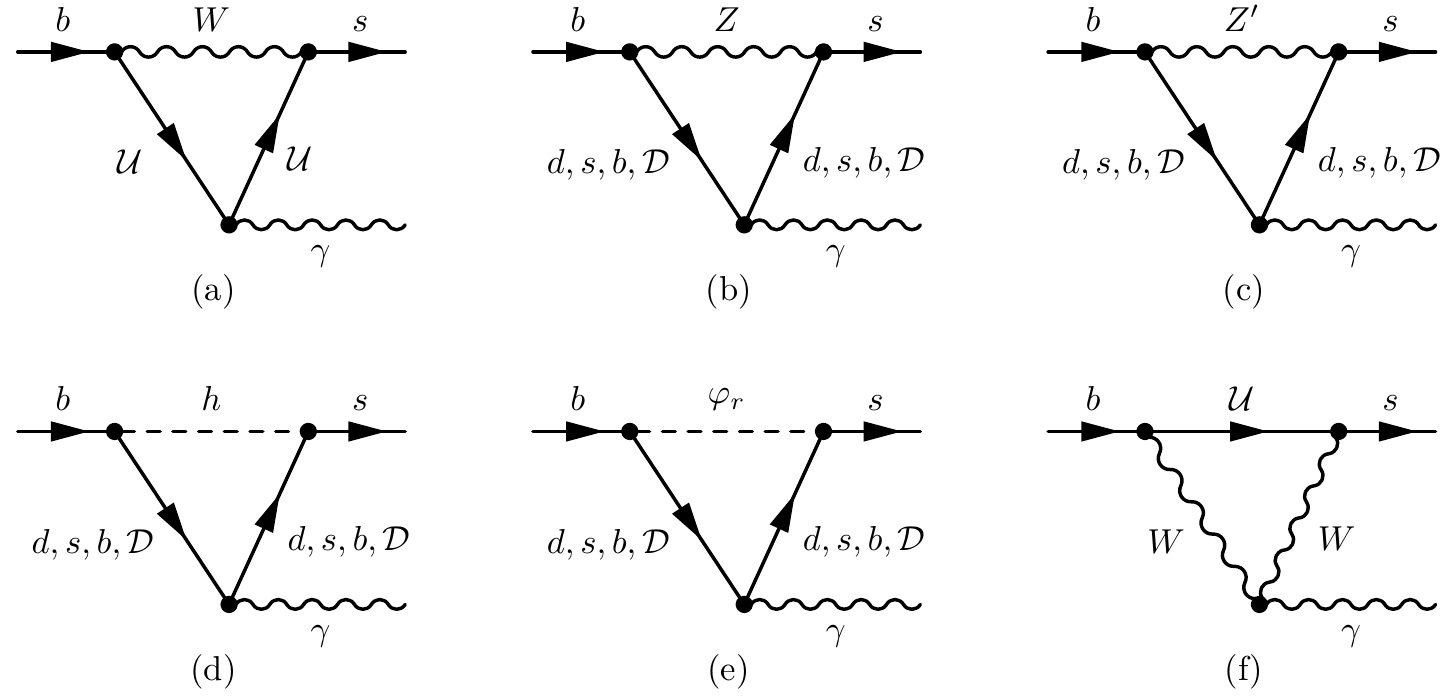}
\caption{Leading new physics contributions to the $b\rightarrow s \gamma$ transition.}
\label{Feynmandiagram}
\end{center}
\end{figure}
%%%%%%%%%%%%%%%%%%%%%%%%%%%%%%%%%%

In the considered model, the Feynman diagrams in the unitary gauge corresponding to the leading new physics contributions to the $b\rightarrow s \gamma$ transition are depicted in Figure \ref{Feynmandiagram}.
We see that beside the contributions of the new particles such as the vectorlike quarks $\mathcal{U}$, $\mathcal{D}$, the $U(1)_X$ gauge boson $Z'$, and the scalar $\varphi_r$, there are also contributions due to the new coupling among the SM particles 
as shown in Figures \ref{Feynmandiagram}b and \ref{Feynmandiagram}d.
These new couplings are induced from the mixings between the SM and the vectorlike quarks \cite{Hieu:2020hti}.

Utilizing the package FeynCalc 
\cite{Mertig:1990an, Shtabovenko:2016sxi, Shtabovenko:2020gxv} for the analytic manipulation of Dirac matrices
 with some further algebraic calculations, the leading new physics contributions to the  Wilson coefficient $ C_7$ have been derived. 
The results are then cross-checked numerically with
 the program Package-X version 2.1.1 
\cite{Patel:2015tea, Patel:2016fam}.
The analytic expression of $C_7^\text{NP}$ is found to be
\begin{eqnarray}
C_7^\text{NP}	
%&=&
%	\frac{ 1}{4\sqrt{2}G_F(V_{tb}V^*_{ts}) m_b}
%	\int_0^1 dx \int_0^{1-x} dy 
%	\left[
%		F_{W\mathcal{U}\mathcal{U}}
%		+ F_{Zqq}
%		+ F_{Z' qq}
%		+ F_{hqq}
%		+ F_{\varphi_r qq}
%		+ F_{WW\mathcal{U}}
%	\right]
%	\nonumber	\\
&=&
	\frac{ 1}{4\sqrt{2}G_F(V_{tb}V^*_{ts}) m_b}
	\int_0^1 dx \int_0^{1-x} dy 
	\left[
		F_{\mathcal{V}ff} 
\Big \vert_{\mathcal{V}=W, \, 		f=\mathcal{U}} +
		\sum_{\mathcal{V}=Z, Z'} \sum_{f=d,s,b,\mathcal{D}} F_{\mathcal{V}ff} 
	\right.
	\nonumber	\\
&&	
	\hspace{7cm}
	\left.
		+ \, F_{WW\mathcal{U}}
		+ \sum_{\mathcal{S}=h, \varphi_r} \sum_{f=d,s,b,\mathcal{D}} F_{\mathcal{S}ff} 
	\right]
\end{eqnarray}
where the loop functions are given by
\begin{eqnarray}
F_{\mathcal{V}ff}	&=&	
	\frac{ Q_f}{m_\mathcal{V}^2}
	\left\lbrace A_1^{\mathcal{V}ff} 
	\Bigl[ 
		(1-3x)
		\ln \bigl[ \beta_{f\mathcal{V}} (x+y)+1-x-y \bigr]
		- x   
	\Bigr] 
	- \frac{A_2^{\mathcal{V}ff}}{\beta_{f\mathcal{V}} (x+y)+1-x-y}   \right\rbrace	
	\nonumber \\		\\
F_{\mathcal{S}ff}	&=&
	\frac{ Q_f}{m_\mathcal{S}^2} \times
	\frac{B^{\mathcal{S}ff}}{\beta_{f\mathcal{S}} (x+y)+1-x-y}	\\
F_{WW\mathcal{U}}	&=&	
	\frac{ 1}{m_W^2}  
	\left\lbrace 
		C_1^{WW\mathcal{U}}  
		\ln \bigl[ (1-\beta_{\mathcal{U}W} ) (x+y)+\beta \bigr]  
		- C_2^{WW\mathcal{U}} 
		+ \frac{C_3^{WW\mathcal{U}} }{(1-\beta_{\mathcal{U}W} ) (x+y)+\beta_{\mathcal{U}W} }   \right\rbrace	
\end{eqnarray}

In the above equations, we have defined 
\begin{eqnarray}
A_1^{\mathcal{V}ff}	&=&	
	g_A^{bf\mathcal{V}} g_A^{sf\mathcal{V}} (2m_f +m_b+m_s) 
	+ g_V^{bf\mathcal{V}} g_V^{sf\mathcal{V}} (-2m_f +m_b+m_s) \nonumber \\
&&	+ \; g_V^{bf\mathcal{V}} g_A^{sf\mathcal{V}} (2m_f -m_b+m_s) 
	+ g_A^{bf\mathcal{V}} g_V^{sf\mathcal{V}} (-2m_f -m_b+m_s)	, \\
A_2^{\mathcal{V}ff}	&=&
	g_A^{bf\mathcal{V}} g_A^{sf\mathcal{V}} 
	\Bigl\{ -4m_f (x+y-1) + (m_b + m_s)\bigl[ \beta_{f\mathcal{V}} (x+y)x + 2(x+y-1)(x-1) \bigr]  
	\Bigr\}  \notag	\\
&&	+\; g_A^{bf\mathcal{V}} g_V^{sf\mathcal{V}} 
	\Bigl\{ 
	4m_f (x+y-1) + (-m_b + m_s) \bigl[ \beta_{f\mathcal{V}} (x+y)x + 2(x+y-1)(x-1) \bigr]  
	\Bigr\}  \notag	\\
&&	+\; g_V^{bf\mathcal{V}} g_A^{sf\mathcal{V}} 
	\Bigl\{ 
	-4m_f (x+y-1) + (-m_b + m_s)\bigl[ \beta_{f\mathcal{V}} (x+y)x + 2(x+y-1)(x-1) \bigr]   
	\Bigr\}  \notag	\\
&&	+\; g_V^{bf\mathcal{V}} g_V^{sf\mathcal{V}} 
	\Bigl\{ 
	4m_f (x+y-1) + (m_b + m_s) \bigl[ \beta_{f\mathcal{V}} (x+y)x + 2(x+y-1)(x-1) \bigr]  
	\Bigr\} ,	\nonumber	\\	
\end{eqnarray}
\begin{eqnarray}
B^{\mathcal{S}ff}	&=&
%	g^{bf\mathcal{S}} g^{sf\mathcal{S}} 
%	\Bigl[ m_f (x+y) - (m_b + m_s)x(x+y-1)  \vphantom{\frac{1}{2}} 
%	\Bigr] ,		\\
	g_P^{bf\mathcal{S}} g_P^{sf\mathcal{S}} 
	\Bigl[ m_f (x+y) + (m_b + m_s)x(x+y-1)  
	\Bigr]  \notag\\
&&	+\; g_P^{bf\mathcal{S}} g_S^{sf\mathcal{S}} 
	\Bigl[ m_f (x+y) - (m_s - m_b)x(x+y-1)  
	\Bigr]  \notag\\
&&	+\; g_S^{bf\mathcal{S}} g_P^{sf\mathcal{S}} 
	\Bigl[ m_f (x+y) + (m_s - m_b)x(x+y-1)  \vphantom{\frac{1}{2}} 
	\Bigr]  \notag\\
&&	+\; g_S^{bf\mathcal{S}} g_S^{sf\mathcal{S}} 
	\Bigl[ m_f (x+y) - (m_b + m_s)x(x+y-1)  \vphantom{\frac{1}{2}} 
	\Bigr] ,		
\end{eqnarray}
\begin{eqnarray}
C_1^{WW\mathcal{U}}	&=&
	g_A^{b\,\mathcal{U}W} g_A^{s\,\mathcal{U}W} 
	\Bigl\{ (m_b + m_s) \bigl[ 4x^2+2x(2y-3)+1 \bigr]  
	\Bigr\}  \notag\\
&&	+\; g_A^{b\,\mathcal{U}W} g_V^{s\,\mathcal{U}W}
	\Bigl\{ (-m_b + m_s) \bigl[ 4x^2+2x(2y-3)+1 \bigr]   
	\Bigr\}  \notag\\
&&	+\; g_V^{b\,\mathcal{U}W} g_A^{s\,\mathcal{U}W} 
	\Bigl\{ (-m_b + m_s) \bigl[4x^2+2x(2y-3)+1 \bigr]  
	\Bigr\}  \notag\\
&&	+\; g_V^{b\,\mathcal{U}W} g_V^{s\,\mathcal{U}W} 
	\Bigl\{ (m_b + m_s) \bigl[4x^2+2x(2y-3)+1 \bigr]  
	\Bigr\} ,	\\
C_2^{WW\mathcal{U}}	&=&
	g_A^{b\,\mathcal{U}W} g_A^{s\,\mathcal{U}W} 
	\Bigl\{ -m_\mathcal{U} - (m_b + m_s) \bigl[ x^2+x(y-2)+1 \bigr] 
	\Bigr\}  \notag\\
&&	+\; g_A^{b\,\mathcal{U}W} g_V^{s\,\mathcal{U}W} 
	\Bigl\{ m_\mathcal{U} + (m_b - m_s) \bigl[x^2+x(y-2)+1 \bigr]  
	\Bigr\}  \notag\\
&&	+\; g_V^{b\,\mathcal{U}W} g_A^{s\,\mathcal{U}W} 
	\Bigl\{ -m_\mathcal{U} + (m_b - m_s) \bigl[x^2+x(y-2)+1 \bigr]   
	\Bigr\}  \notag\\
&&	+\; g_V^{b\,\mathcal{U}W} g_V^{s\,\mathcal{U}W} 
	\Bigl\{ m_\mathcal{U} - (m_b + m_s) \bigl[x^2+x(y-2)+1 \bigr]  
	\Bigr\} ,	\\
%
%
%\end{eqnarray}
%\begin{eqnarray}
C_3^{WW\mathcal{U}}	&=&
	g_A^{b\,\mathcal{U}W} g_A^{s\,\mathcal{U}W} 
	\Bigl[ 3m_\mathcal{U} (x+y) + (m_b + m_s)(2x^2+2xy+y)  \Bigr] \notag\\
&&		
	+\; g_A^{b\,\mathcal{U}W} g_V^{s\,\mathcal{U}W} 
	\Bigl[ -3m_\mathcal{U} (x+y) + (-m_b + m_s)(2x^2+2xy+y)  \Bigr]  \notag\\
&&		
	+ \; g_V^{b\,\mathcal{U}W} g_A^{s\,\mathcal{U}W} 
	\Bigl[ 3m_\mathcal{U} (x+y) +\; (-m_b + m_s)(2x^2+2xy+y)  \Bigr]  \notag\\
&&		
	+\; g_V^{b\,\mathcal{U}W} g_V^{s\,\mathcal{U}W} 
	\Bigl[ -3m_\mathcal{U} (x+y) + (m_b + m_s)(2x^2+2xy+y)  \Bigr] ,
\end{eqnarray}
and $\beta_{ab} =\dfrac{m_a^2}{m_b^2}$.
In these above equations, $g_V$ and $g_A$ are the vector and axial-vector couplings between fermions and a gauge boson,
while $g_S$ and $g_P$ are the scalar and pseudo-scalar couplings between fermions and a scalar field.

%
%===================
%SM contribution is at NNLO,

%%%%%%%%%%%%%%%%%%%%%%%%%%%%%%%
\section{Numerical analysis}
%%%%%%%%%%%%%%%%%%%%%%%%%%%%%%%

In the numerical analysis, we start with a quark basis where the $3 \times 3$ blocks of the up-type and down-type quark mass matrices are diagonal with the values being the center experimental values of the SM quark masses.
In this basis, the full $4 \times 4$ quark mass matrices are represented as
\begin{eqnarray}
M^u	=
	\begin{pmatrix}
	m_u^0	&	0	&	0	&	w_1 \langle \phi \rangle	\\
	0	&	m_c^0	&	0	&	w_2 \langle \phi \rangle	\\
	0	&	0	&	m_t^0	&	w_3 \langle \phi \rangle	\\
	0	&	0	&	0	&	m_U
	\end{pmatrix} ,
	\quad
M^d	=
	\begin{pmatrix}
	m_d^0	&	0	&	0	&	w_1 \langle \phi \rangle	\\
	0	&	m_s^0	&	0	&	w_2 \langle \phi \rangle	\\
	0	&	0	&	m_b^0	&	w_3 \langle \phi \rangle	\\
	0	&	0	&	0	&	m_D
	\end{pmatrix} ,
\end{eqnarray}
Here, for simplicity, we consider the case of degenerate up-type and down-type vectorlike fermion masses, namely
$m_U = m_D = m_Q$ and
$m_N = m_E = m_L$.
After the diagonalization of these mass matrices, the masses of the SM quarks are deflected from their center values due to the mixing.
They are required to stay within the $2\sigma$ ranges of the experimental measurements.

In this model, the SM CKM matrix is a $3 \times 3$ block of a $4 \times 4$ matrix determined by 
$\left( V_L^u V_L^{d\dagger} \right)_{4\times 4}$.
Due to the mixing between the SM and the vectorlike quarks, the observed unitarity violation of the SM CKM matrix can be explained.
Here, we consider the experimental constraint on the sum of the squared modules of the first-row elements of the SM CKM matrix \cite{Workman:2022ynf}:
\begin{eqnarray}
0.9971
\leq |V_{ud}|^2 + |V_{us}|^2 + |V_{ub}|^2 \leq
0.9999 \quad (2 \sigma).
\label{ckm}
\end{eqnarray}

In our numerical analysis, 
we consider the current 2$\sigma$ range of the branching ratio of radiative $B$-meson decay \cite{HFLAV:2022pwe}:
\begin{eqnarray}
3.11 \times 10^{-4} < \text{BR}(B \rightarrow X_s \gamma) < 3.87 \times 10^{-4}.
\label{bsg_now}
\end{eqnarray}
Assuming the center value of this branching ratio remains unchanged, we can expect that the relative precision can reach 2\% with the luminosity 50 ab$^{-1}$ at the Belle II experiment
\cite{Belle-II:2022cgf, DiCanto:2022icc}.
The impact of the future results is investigated by taking into account the following expected 2$\sigma$ range:
\begin{eqnarray}
3.35 \times 10^{-4} < \text{BR}(B \rightarrow X_s \gamma)_{\text{Belle2}} < 3.63 \times 10^{-4}.
\label{bsg_belle2}
\end{eqnarray}

The constraints from semileptonic $B$-meson decays are also considered. They are the branching ratios of the processes 
$B^+ \rightarrow K^+ \mu^+ \mu^-$ 
\cite{BR, Aaij:2012vr, Aaij:2014pli},
$B^0 \rightarrow K^*(892)^{0} \mu^+ \mu^-$ 
\cite{BR, Aaij:2016flj},
$B^+ \rightarrow K^+ e^+ e^-$
\cite{LHCb:2022qnv}, and
$B^0 \rightarrow K^*(892)^{0} e^+ e^-$
\cite{LHCb:2022qnv}.
The theoretical predictions of these branching fractions are calculated using the Wilson coefficients $C^{(')}_{9,10}$, to which the new physics contributions are given in Appendix \ref{C9-10}.
%
%and the observables
%$R_K$ \cite{HFLAV:2022pwe, Aaij:2019wad, Aaij:2014ora, Aaij:2021vac, LHCb:2022qnv}
%and 
%$R_{K^*}$ \cite{HFLAV:2022pwe, Aaij:2017vbb, Abdesselam:2019wac, LHCb:2022qnv}
%measuring the lepton flavor violation.
%
The current 2$\sigma$ allowed ranges corresponding to these quantities for the lepton invariant mass in the region 
$q^2 = [1.1, 6.0]$ GeV$^2$
are given as follows
\begin{eqnarray}
1.050 \times 10^{-7} < \text{BR}(B^+ \rightarrow K^+ \mu^+ \mu^-) < 1.322 \times 10^{-7} , 
\label{BRK}		\\
1.382 \times 10^{-7} < \text{BR}(B^0 \rightarrow K^*(892)^{0} \mu^+ \mu^-) < 1.970 \times 10^{-7} ,
\label{BRKs}		\\
1.090 \times 10^{-7} < \text{BR}(B^+ \rightarrow K^+ e^+ e^-) < 1.416 \times 10^{-7}, 
\label{BRKe}
\\
1.290 \times 10^{-7} < \text{BR}(B^0 \rightarrow K^*(892)^0 e^+ e^-) < 1.965 \times 10^{-7}. 
\label{BRKse}
\end{eqnarray}

%\begin{eqnarray}
%\\
%%0.762 
%\textcolor{blue}{0.855} < R_K = \frac{\text{BR}(B^+ \rightarrow K^+ \mu^+ \mu^-)}{\text{BR}(B^+ \rightarrow K^+ e^+ e^-)} < 
%\textcolor{blue}{1.043}
%%0.930 
%, 
%\label{RK}	\\
%%
%%0.54
%\textcolor{blue}{0.877} < R_{K ^*} = \frac{\text{BR}(B^0 \rightarrow K^*(892)^{0} \mu^+ \mu^-)}{\text{BR}(B^0 \rightarrow K^*(892)^{0} e^+ e^-)} < 
%\textcolor{blue}{1.177}
%%0.96 
%.
%\label{RKs}
%\end{eqnarray}

The slepton searches at the ATLAS and CMS experiments at 13 GeV impose the constraints on the charged vectorlike lepton masses 
\cite{Aad:2014vma}
that satisfy either 
$m_L \gtrsim \mathcal{O}(1)$ TeV, or 
\begin{eqnarray}
m_L - m_{\chi_r} \lesssim 60 \text{ GeV}.
\label{vectorlikelepton}
\end{eqnarray}
On the other hand, in order to explain the deviation between the current experimental value \cite{Muong-2:2023cdq} 
and the SM prediction of the muon anomalous magnetic moment
\cite{Whitepaper2020, Tanabashi:2018oca, Abi:2021gix, Aoyama:2020ynm}
\begin{eqnarray}
\Delta a_\mu \equiv 
	a_\mu^\text{exp} - a_\mu^\text{SM}
	= (24.9 \pm 4.8) \times 10^{-10} \, ,
\label{g-2}
\end{eqnarray}
the vectorlike leptons must be light enough.
%

%============================

%+ Constraints on $m_{Z'}$: 

In our considered model, the $Z'$ boson does not couple to the SM leptons and the SM Higgs boson, while the gauge kinetic mixing is zero. 
The couplings between this gauge boson and the SM quarks are generated by the mixing between the SM quarks and the vectorlike quarks that are severely restricted by the FCNC constraints.
We have estimated that the cross sections in the searches for the $Z'$ boson are indeed negligible compared to the corresponding SM backgrounds.
Therefore, the parameter regions compatible with the FCNC constraints satisfies the current collider bounds imposed by the $Z'$-boson searches 
\cite{Lees:2014xha, Anastasi:2018azp, 
ATLAS:2019erb, Sirunyan:2019wqq, 
CMS:2021sch,
Aaij:2019bvg, ATLAS:2020lks,
Belle-II:2022jyy, ATLAS:2023jyp}.

%+ Constraints on vectorlike quarks

At the LHC, the vectorlike quarks can be produced either singly or in pairs.
In the single-production searches
\cite{ATLAS:2023pja, ATLAS:2023bfh, 
CMS:2022yxp}, 
the constraints are set on the parameter space of the vectorlike quark mass and the universal coupling strengths of the electroweak/Yukawa interactions between the vectorlike quarks, the SM quarks and the SM massive gauge/Higgs bosons.
Since these couplings strengths are determined by the mixing matrices between the vectorlike quarks and the SM ones, they are strongly suppressed due to the stringent requirements on the FCNCs and the CKM unitarity.
As a result, once the constraints on the FCNCs and the CKM unitarity are imposed, the viable parameter space of the model also satisfies the current constraints from the single-production searches by the ATLAS and the CMS experiments at the LHC.
For the pair-production searches \cite{ATLAS:2018ziw, ATLAS:2022hnn, 
ATLAS:2022tla, 
CMS:2018zkf, CMS:2020ttz}, 
the vectorlike quarks are created mainly via the strong interactions with the cross section depends only on the vectorlike quark mass. 
This mode is dominant for the range of the vectorlike quark masses up to about $\mathcal{O}(1)$ TeV.
Hence, the vectorlike quark mass in our model is subjected to the constraint from the pair-production searches at the LHC, for which the current most severe lower bound is
\cite{ATLAS:2022tla}
\begin{eqnarray}
m_Q \gtrsim 1.59 \text{ TeV.}
\label{mQ_limit}
\end{eqnarray}

%============================

The set of free inputs of the model includes
\begin{eqnarray}
y_{1,2,3},
g_X,
w_{1,2,3},
m_{Z'},
\lambda_\phi,
m_Q,
m_{\chi_r},
\tau = \frac{m_L^2}{m_{\chi_r}^2},
\delta=\frac{m_{\chi_i}^2}{m_{\chi_r}^2} - 1.
\end{eqnarray}
Noticing that the new physics contribution to the muon anomalous magnetic moment%
\footnote{
The formula of the new physics contribution to the muon $g-2$ is given in Appendix \ref{appendix_g-2}.
} 
depends on the parameters $y_2$, 
$m_{\chi_r}$,
$\tau = \frac{m_L^2}{m_{\chi_r}^2}$, and 
$\delta= \frac{m_L^2}{m_{\chi_r}^2}$,
to reconcile the two constraints in
Eqs. (\ref{vectorlikelepton}) and (\ref{g-2}),
we choose these parameters to be
%
%$y_2 = 3$
$y_2 = 3.3$,
$m_{\chi_r} = 120$ GeV,
$\tau = 1.78$, and
$\delta = 1$
\cite{Dinh:2020inx}.
To explain the large deviations between the SM prediction and the experimental value of 
BR($B^+ \rightarrow K^+ \mu^+ \mu^-$), the large exotic Yukawa coupling $y_{2}$ are essential.
Meanwhile, the sizable $y_1$ plays an important role in explaining the enhancement in the electron channels of the semileptonic B-meson decay announced recently by the LHCb Collaboration \cite{LHCb:2022qnv}.
Here, we consider the benchmark case with $y_1 = 3$.
% whenever it is relevant and not mentioned.
In our analysis, the parameter $y_{3}$ is set to zero for simplicity.
In order to suppress the new physics contributions to the FCNCs related to the first two generations, we consider the case of vanishing $w_1$.
Since the effects of $\lambda_\phi$ on the considered observables are negligible, we take $\lambda_\phi = 3$ in our numerical analysis as an example without lost of generality.
The remaining set of inputs are
$g_X$,
$w_2$, 
$w_3$,
$m_{Z'}$, and
$m_Q$.
For the $U(1)_X$ gauge couplings and the exotic Yukawa couplings, we further impose the perturbation limits as
\begin{eqnarray}
g_X,\, w_{2,3} \leqslant \sqrt{4\pi}.
\label{perturbation}
\end{eqnarray}

%==============================
%
%Constraint from LHC in searches for leptons, scalars, vectorlike quarks (~1-1.5 TeV: 
%2212.05263,
%2210.15413,
%2201.07045,
%1808.01771, 
%1806.10555, 
%1509.04261), vectorlike leptons
%
%=====================
%
%Explain about the choice of basis so that the 3x3 block of the mass matrix and 3x3 block of the CKM matrix is close to those the SM 
%$=>$ this is used for convenience in numerical analysis.
%
%==================================

%%%%%%%%%%%%%%%%%%%%%%%%%%%%%%%%%%
\begin{figure}[h!]
\begin{center}
\includegraphics[scale=0.65]{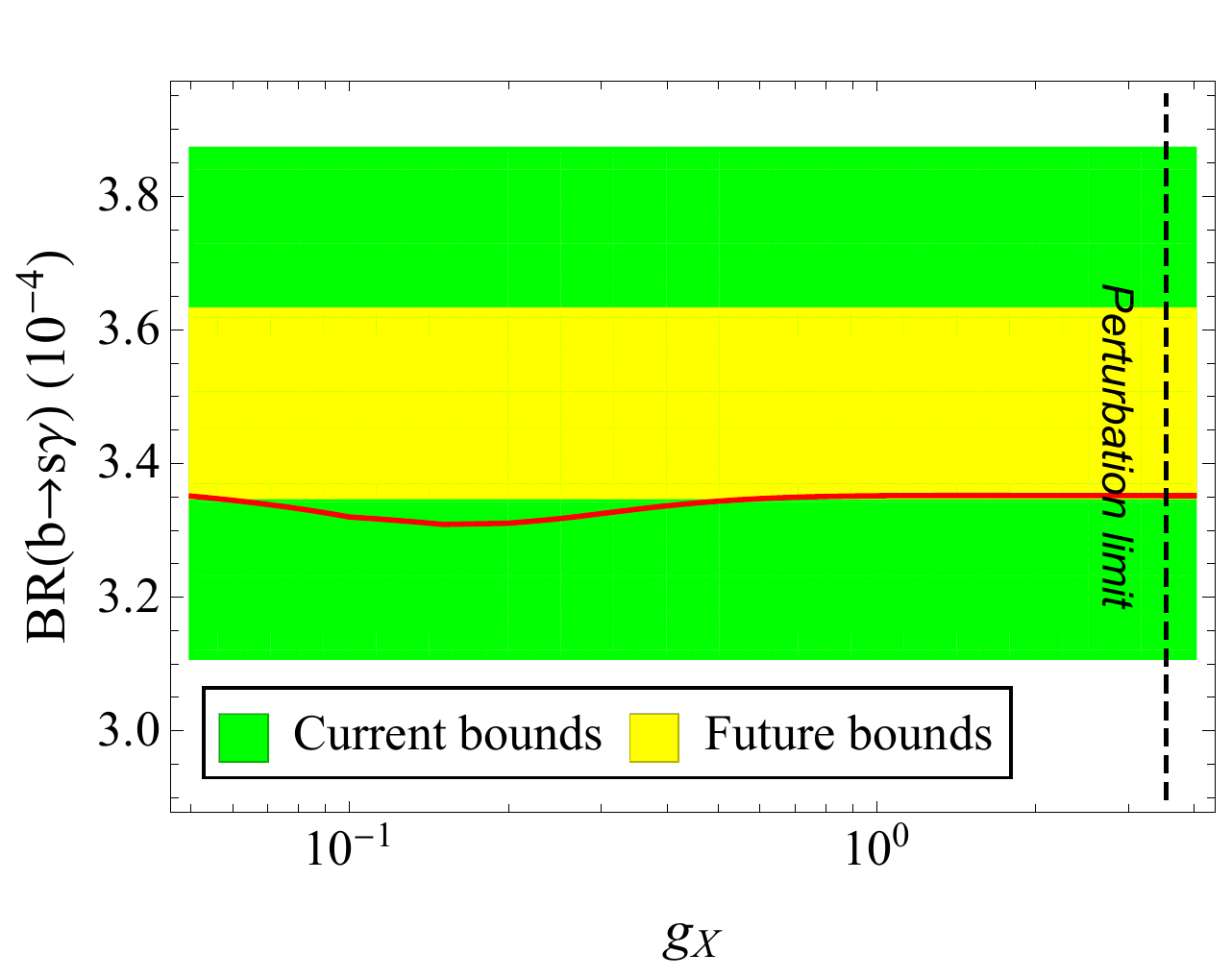}
\caption{The branching ratio of the $b\rightarrow s \gamma$ transition as a function of $g_X$ when
$m_{Z'}=700$ GeV,
$m_Q = 2500$ GeV, 
%$\lambda_\phi = 3$,
%$w_1 = 0$,
$w_2 = 2.45$, 
$w_3 = 0.25$.
The green and yellow regions indicate the current bounds and expected ones at the Belle II experiment in the near future.
The vertical black dashed line indicates the perturbation limit (\ref{perturbation}).}
\label{BRgx}
\end{center}
\end{figure}
%%%%%%%%%%%%%%%%%%%%%%%%%%%%%%%%%%

The branching ratio of the $b\rightarrow s \gamma$ process is calculated according to the method in Refs. 
\cite{Misiak:2006ab, Czakon:2015exa, Misiak:2020vlo, HFLAV:2022pwe}.
In Figure \ref{BRgx}, this branching ratio is plotted as a function of $g_X$ in the benchmark case with 
$m_{Z'}=700$ GeV,
$m_Q = 2500$ GeV, 
%$\lambda_\phi = 3$,
%$w_1 = 0$,
$w_2 = 2.45$, 
$w_3 = 0.25$.
For this parameter setting, 
the most dominant contribution to this process come from the loop involving the $U(1)_X$ gauge boson $Z'$ in Figures \ref{Feynmandiagram}c.
Note that the coupling of $Z'$ and the down-type quark current proportional to the product of $g_X$ and the mixing matrices that, in turn, is roughly proportional to $g_X \langle \phi \rangle$.
Therefore, according to Eq. (\ref{mZp}), once $m_{Z'}$ is fixed, this coupling does not depend on $g_X$.
This explains the behavior of the branching ratio in Figure \ref{BRgx}.
This benchmark satisfies the current bounds (\ref{bsg_now}) depicted by the green region for the whole plotted range of $g_X$ upto the perturbation limit (the vertical black dashed line).
In the near future, if the center value of the branching ratio remains unchanged,
the 2$\sigma$ allowed region (\ref{bsg_belle2}) expected at the Belle II experiment is depicted in this figure as the yellow band.
The expected lower bound will be able to exclude a certain range of $g_X$ between 0.05 and 0.79 in this case.

%%%%%%%%%%%%%%%%%%%%%%%%%%%%%%%%%%
\begin{figure}[h!]
\begin{center}
\includegraphics[scale=0.65]{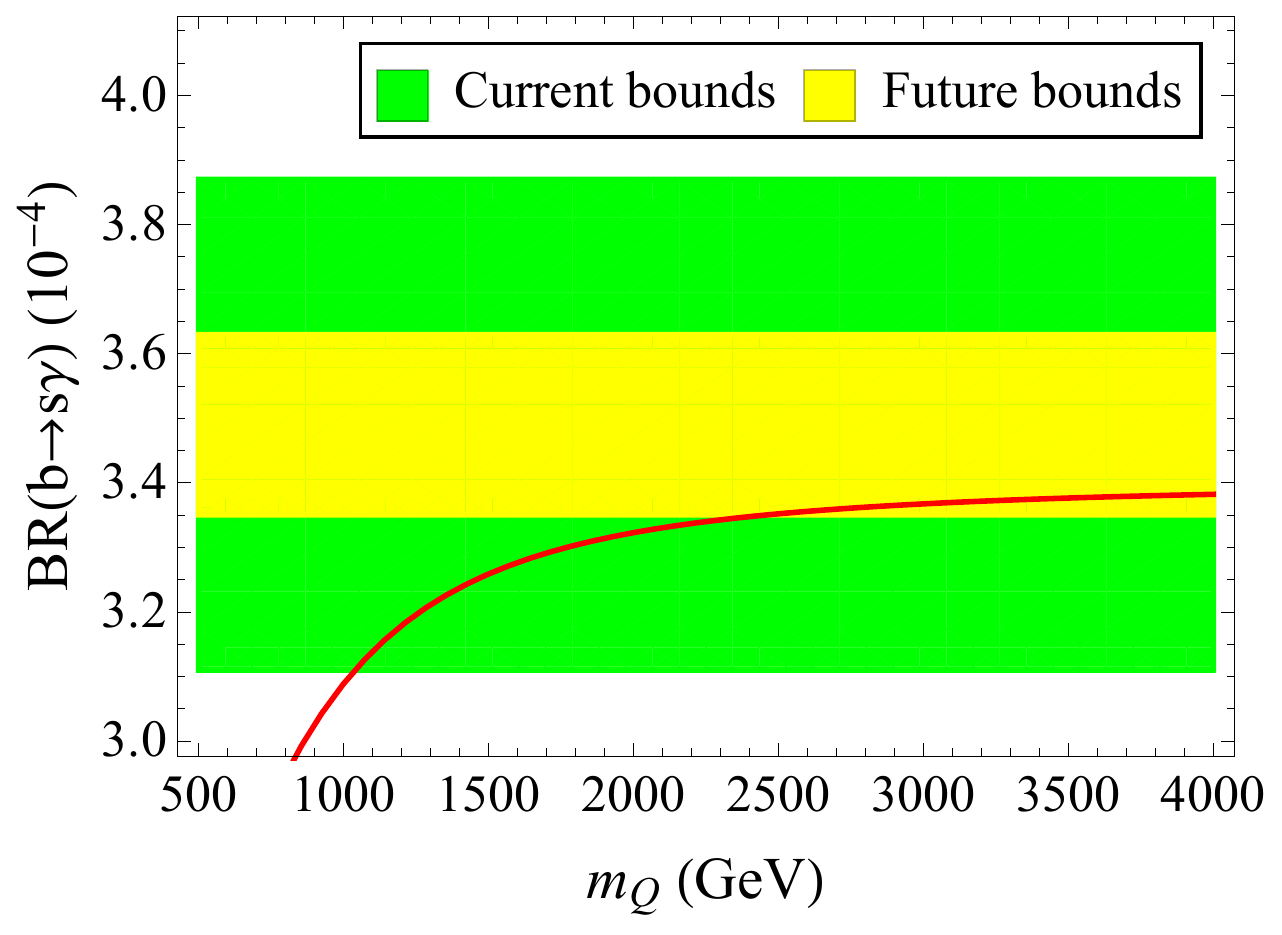}
\caption{The branching ratio of the $b\rightarrow s \gamma$ transition as a function of $m_Q$ when
$m_{Z'}=700$ GeV,
$g_X = 1.4$ GeV, 
%$\lambda_\phi = 3$,
%$w_1 = 0$,
$w_2 = 2.45$, 
$w_3 = 0.25$.
The color codes are the same as those in Figure \ref{BRgx}.}
\label{BRm4}
\end{center}
\end{figure}
%%%%%%%%%%%%%%%%%%%%%%%%%%%%%%%%%%

In Figure \ref{BRm4}, $\text{BR}(b \rightarrow s \gamma)$ is shown as a function of the vectorlike quark mass $m_Q$ for the case of 
$g_X = 1.4$, 
while other parameters are the same as those in Figure \ref{BRgx}.
We observe that the branching ratio can be significantly enhanced by larger values of the vectorlike quark mass $m_Q$.
As $m_Q$ increases, the vector-like quarks gradually decouple from the SM sector at low energies.
Therefore, the branching ratio in this model approaches the SM limit for large values of $m_Q$.
The current $2\sigma$ bounds (\ref{bsg_now}) of this branching ratio (the green region) set the lower limit on $m_Q$ that is approximately 
1050 GeV.
After getting the results from the Belle II experiment, the expected 2$\sigma$ bounds in Eq. (\ref{bsg_belle2}) (the yellow region) will raise the lower limit of $m_Q$ up to about 
2360 GeV.

%%%%%%%%%%%%%%%%%%%%%%%%%%%%%%%%%%
\begin{figure}[h!]
\begin{center}
\includegraphics[scale=0.65]{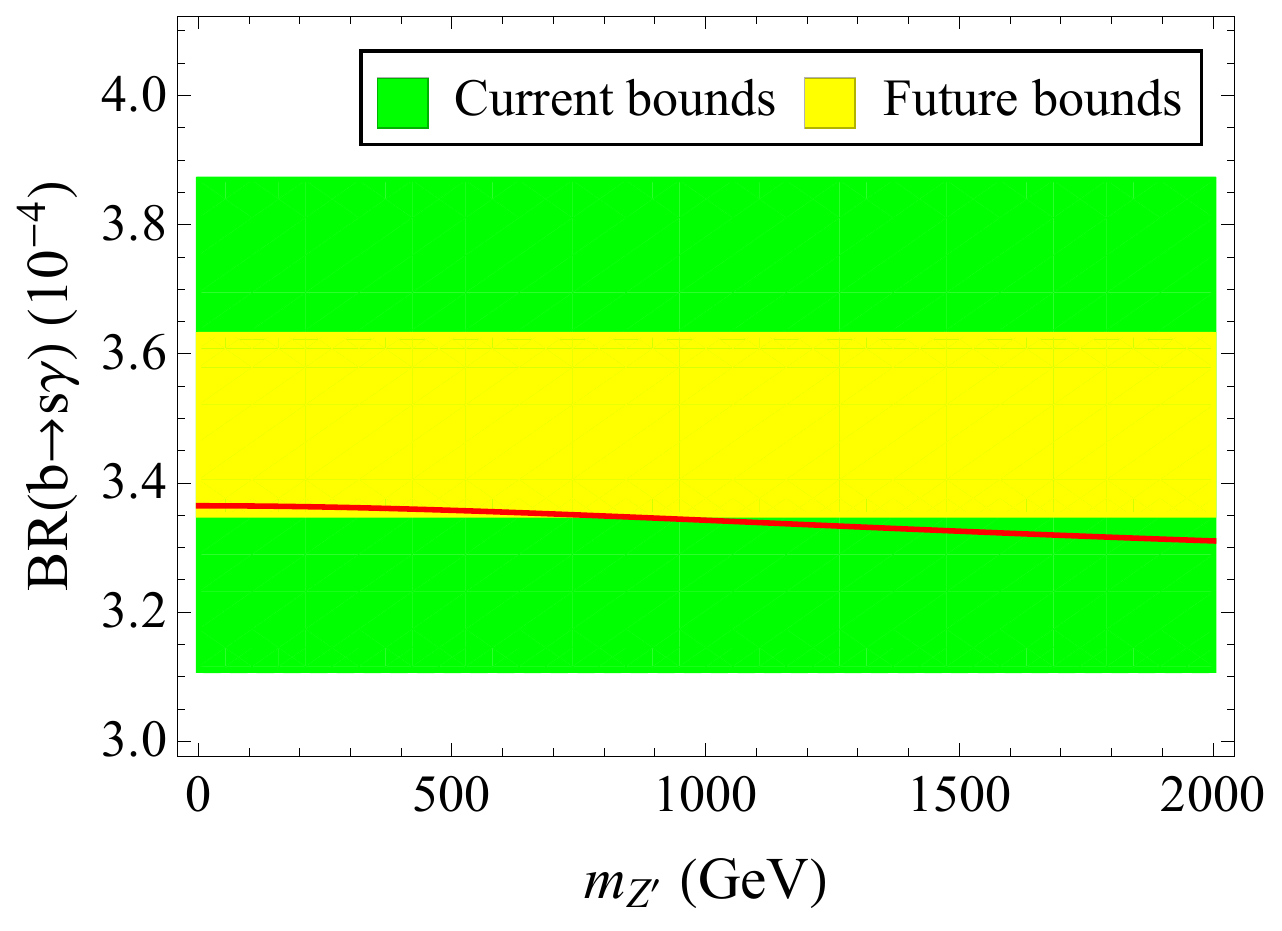}
\caption{The branching ratio of the $b\rightarrow s \gamma$ transition as a function of $m_{Z'}$ when
$m_{Q}=2500$ GeV,
$g_X = 1.4$, 
%$\lambda_\phi = 3$,
%$w_1 = 0$,
$w_2 = 2.45$, 
$w_3 = 0.25$.
The color codes are the same as those in Figure \ref{BRgx}.}
\label{BRmzp}
\end{center}
\end{figure}
%%%%%%%%%%%%%%%%%%%%%%%%%%%%%%%%%%

The dependence of $\text{BR}(b\rightarrow s\gamma)$ on $m_{Z'}$ is depicted in Figure \ref{BRmzp} for the case with 
$g_X = 1.4$ 
and other parameters are the same as those in Figure \ref{BRgx}.
We observe that the transition rate is slightly reduced when $m_{Z'}$ is increased.
Since the main contribution comes from the diagram in Figure \ref{Feynmandiagram}c, this dependence is not strong for a fixed value of $g_X$.
% because there is a compensation between the dependences of the squared coupling and Z' propagator on $m_{Z'}$ in the loop diagram of Figure 1c.
%
The branching ratio satisfies the current 2$\sigma$ bounds (\ref{bsg_now}) (the green region) for the whole considered range of $m_{Z'}$.
However, it is expected that the constraint (\ref{bsg_belle2}) from the future Belle II result (the yellow region) will set a severe upper limit on the $Z'$ boson mass to be about 
750 GeV for this benchmark case.

%%%%%%%%%%%%%%%%%%%%%%%%%%%%%%%%%%
\begin{figure}[h!]
\begin{center}
\includegraphics[scale=0.65]{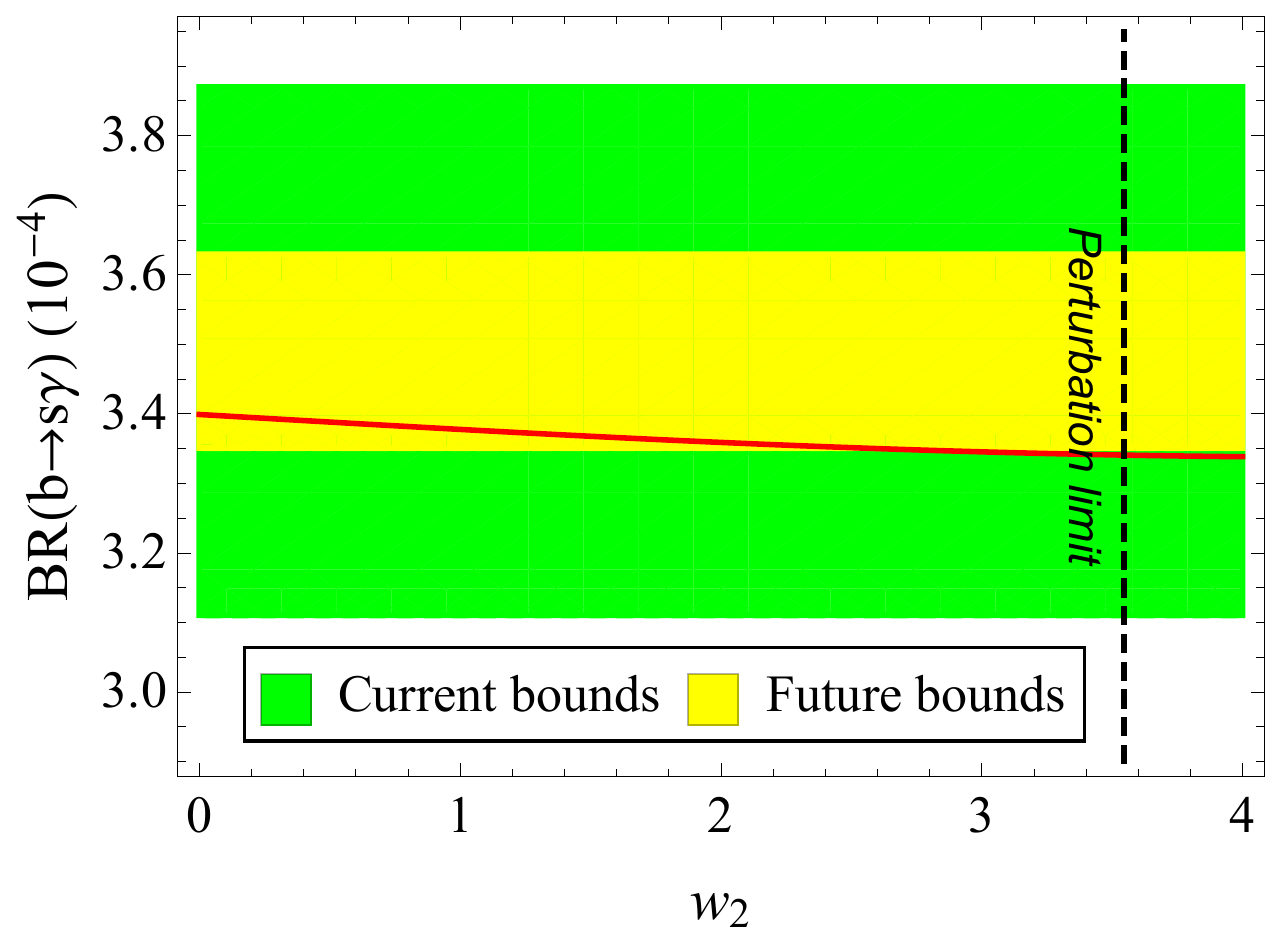}
\caption{The branching ratio of the $b\rightarrow s \gamma$ transition as a function of $w_2$ when
$m_{Q}=2500$ GeV,
$m_{Z'} = 700$ GeV,
$g_X = 1.4$, 
%$\lambda_\phi = 3$,
%$w_1 = 0$,
%$w_2 = 1.55$, 
$w_3 = 0.25$.
The color codes are the same as those in Figure \ref{BRgx}.}
\label{BRw2}
\end{center}
\end{figure}
%%%%%%%%%%%%%%%%%%%%%%%%%%%%%%%%%%

%%%%%%%%%%%%%%%%%%%%%%%%%%%%%%%%%%
\begin{figure}[h!]
\begin{center}
\includegraphics[scale=0.65]{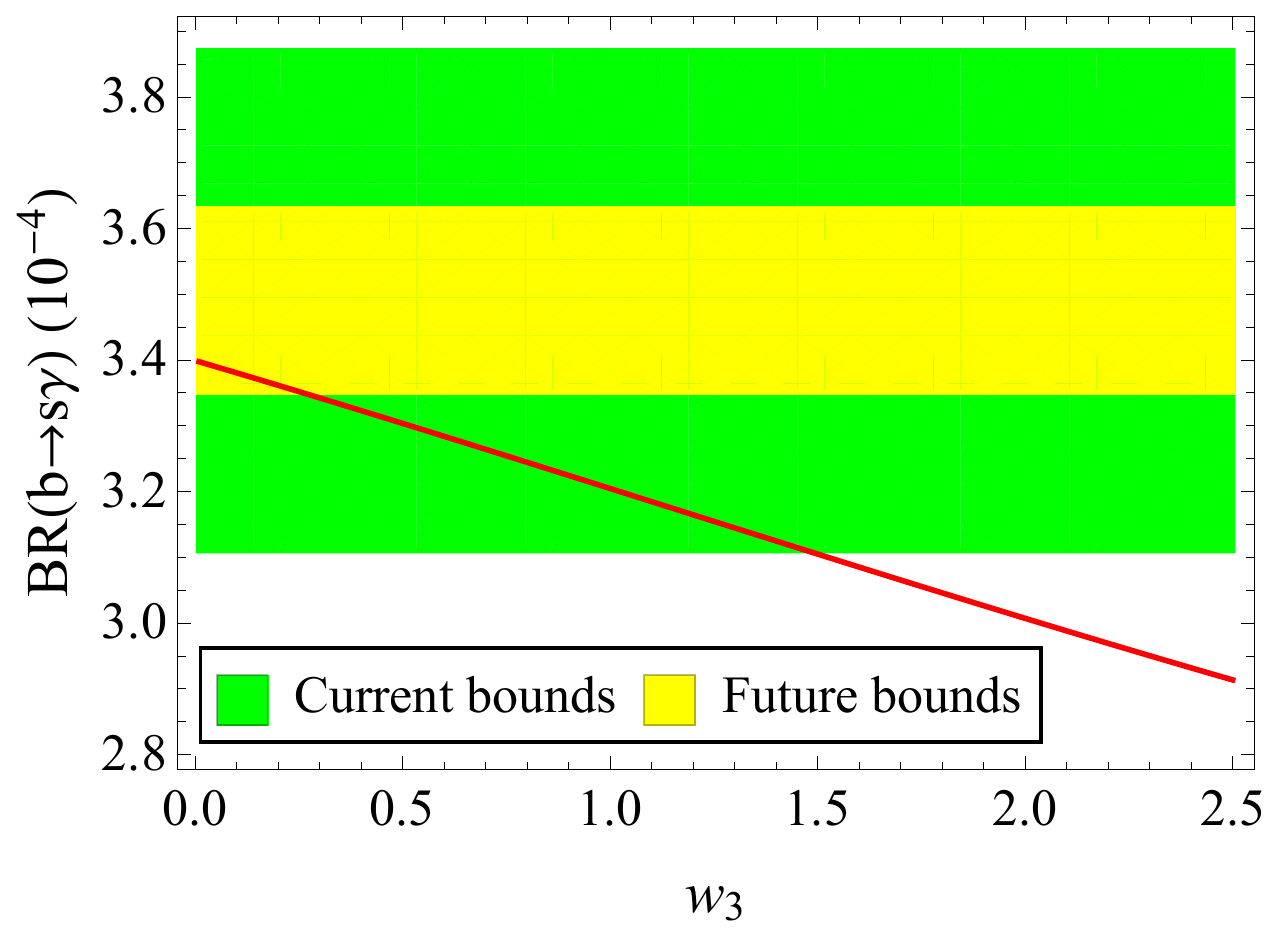}
\caption{The branching ratio of the $b\rightarrow s \gamma$ transition as a function of $w_3$ when
$m_{Q}=2500$ GeV,
$m_{Z'} = 700$ GeV,
$g_X = 1.4$, 
%$\lambda_\phi = 3$,
%$w_1 = 0$,
$w_2 = 2.45$. 
%$w_3 = 0.145$.
The color codes are the same as those in Figure \ref{BRgx}.}
\label{BRw3}
\end{center}
\end{figure}
%%%%%%%%%%%%%%%%%%%%%%%%%%%%%%%%%%

Figure \ref{BRw2} shows the dependence of BR$(b \rightarrow s \gamma)$ on the parameter $w_2$ in the case with 
$m_{Q}=2500$ GeV,
$m_{Z'} = 700$ GeV,
$g_X = 1.4$, 
%$\lambda_\phi = 3$,
%$w_1 = 0$,
%$w_2 = 1.55$, 
$w_3 = 0.25$.
It is observed that the branching ratio is slightly reduced when $|w_2|$ increases.
For this benchmark, the branching ratio stays in the green region allowed by the current constraint (\ref{bsg_now}) on the $b \rightarrow s \gamma$ transition for the whole range of $w_2$ upto the perturbation limit.
In the near future, the Belle II result (\ref{bsg_belle2}) is expected to set the upper limit of about 
2.57 on the parameter $w_2$ 
that is 
below the perturbation limit (\ref{perturbation}).
The branching ratio of the process $b\rightarrow s \gamma$ is plotted as a function of the parameter $w_3$ in Figure \ref{BRw3}.
Similar to Figure \ref{BRw2}, here we observe that the branching ratio is also inversely proportional to $w_3$.
The current constraint (\ref{bsg_now}) requires that $w_3$ must be smaller than 1.5.
Due to the strong dependence of BR$(b\rightarrow s \gamma)$ on $w_3$, this upper bound is expected to be reduced to about 0.27
by the foreseen constraint (\ref{bsg_belle2}) after the Belle II experiment accumulates enough data.

%==============================
%Combined rare $B$ decay constraints
%==============================

%==============
%Plot of RK, RK*, BR(B to K mumu), BR(B to K* mumu)
%==============

%%%%%%%%%%%%%%%%%%%%%%%%%%%%%%%%%%
\begin{figure}[h!]
\begin{center}
\includegraphics[scale=0.65]{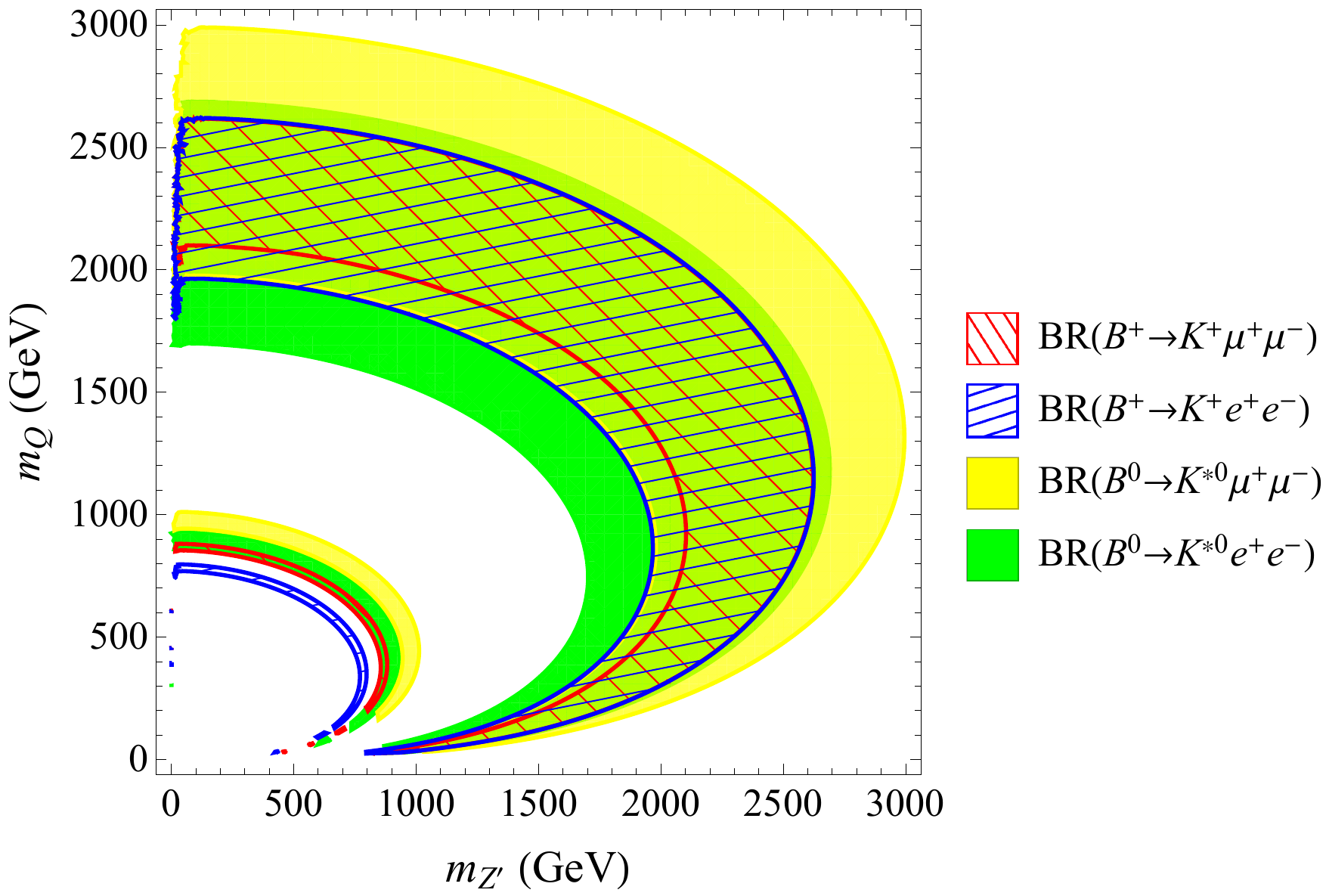}
\caption{Semileptonic decay constraints
(Eqs. (\ref{BRK})-(\ref{BRKse})) at the level of 2$\sigma$ on the 
$(m_{Z'}, m_Q)$ plane
 when
%$m_{Z'}=700$ GeV,
$g_X = 1.4$, 
%$\lambda_\phi = 3$,
%$w_1 = 0$,
$w_2 = 2.45$, 
$w_3 = 0.25$.
%$y_1 = 3.0$,
%$y_2 = 3.3$.
}
\label{bsllmzpmq}
\end{center}
\end{figure}
%%%%%%%%%%%%%%%%%%%%%%%%%%%%%%%%%%

In Figure \ref{bsllmzpmq}, the constraints 
on the semileptontic decays of $B$ mesons are plotted on the ($m_{Z'}, m_Q$) plane for the case with
$g_X = 1.4$, 
%$\lambda_\phi = 3$, 
$w_2 = 2.45$, and
%$w_1 = 0$,
%$w_2 = 1.55$, and
$w_3 = 0.25$.
%$y_1 = 3.0$, 
%$y_2 = 3.3$.
The 2$\sigma$ allowed regions corresponding to the constraints 
%(Eqs. (\ref{BRK})-(\ref{BRKs})) 
on 
BR$(B^+ \rightarrow K^+ \mu^+ \mu^-)$
(\ref{BRK}),
%$R_K$ (\ref{RK}), 
BR($B^0 \rightarrow K^{0*} \mu^+ \mu^-$)
(\ref{BRKs}), 
%and $R_{K^*}$ (\ref{RKs})
BR($B^+ \rightarrow K^+ e^+ e^-$) (\ref{BRKe}), and
BR($B^0 \rightarrow K^{0*} e^+ e^-$) (\ref{BRKse})
are shown as 
the red back-hatched, 
the yellow, 
the blue hatched and 
the green areas.
Each constraint has two allowed regions, the narrow one corresponding to small values of both $m_{Z'}$ and $m_Q$, 
and
the wider region including points with larger values of these two parameters.
From this figure, we see that only the latter has an overlapping region
satisfying all the above four constraints.
The boundaries of this allowed region are formed by the constraints on the branching fractions of the decay processes
$B^+ \rightarrow K^+ \mu^+ \mu^-$ and
$B^+ \rightarrow K^+ e^+ e^-$, of which the large deviations from the SM predictions are observed.

%==============================
%Combined rare $B$ decay constraints
%==============================

%%%%%%%%%%%%%%%%%%%%%%%%%%%%%%%%%%
\begin{figure}[h!]
\begin{center}
\includegraphics[scale=0.65]{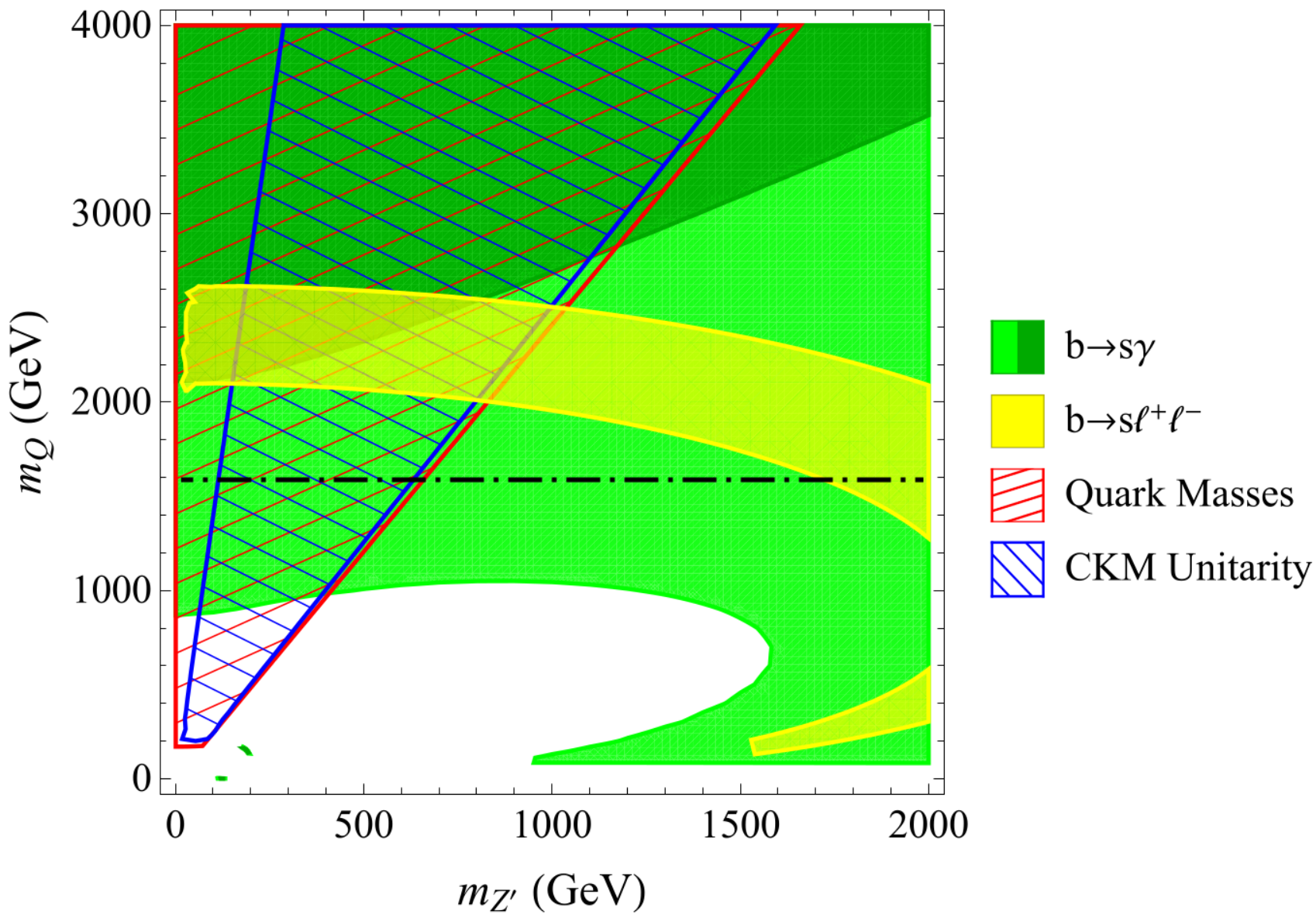}
\caption{Constraints on the $(m_{Z'},m_Q)$ plane
 when
%$m_{Q}=1700$ GeV,
$g_X = 1.4$, 
%$\lambda_\phi = 3$,
%$w_1 = 0$,
$w_2 = 2.45$, 
$w_3 = 0.25$.
The light and the dark green colors indicate the current 2$\sigma$ allowed region (\ref{bsg_now})
and the expected one after the Belle II experiment (\ref{bsg_belle2}).
The constraints on the SM quark masses are shown by the red hatched region.
The blue back-hatched region satisfies the constraint  (\ref{ckm}) on the CKM unitarity violation of the first row.
The horizontal dash-dotted line indicates the lower bound (\ref{mQ_limit}) on $m_Q$.}
\label{mzpmq}
\end{center}
\end{figure}
%%%%%%%%%%%%%%%%%%%%%%%%%%%%%%%%%%

The above overlapping 
region is extracted and plotted in Figure \ref{mzpmq} as a yellow band.
In addition, we consider other constraints on the $b\rightarrow s \gamma$ transition (\ref{bsg_now}),
the SM quark masses, 
the violation of the CKM unitarity in the first row (\ref{ckm}), 
and the vectorlike quark mass (\ref{mQ_limit}).
%
%The constraints on the plane ($m_{Z'}, m_Q$) are plotted in Figure \ref{mzpmq} for the case with
%$g_X = 2$, 
%$\lambda_\phi = 3$,
%$w_1 = 0$,
%$w_2 = 1.55$, and
%$w_3 = 0.145$.
%
In this figure, the CKM unitarity violation (the blue back-hatched region) imposes the allowed range on the mass ratio $\frac{m_Q}{m_{Z'}}$ to be 
[2.5, 13.8].
For a given value of $m_Q$, the region with too large $m_{Z'}$ generates large mixing between the vectorlike and the SM quarks causing the CKM to be too far away from unitarity. Hence, it is excluded.
On the other hand, the region with too small $m_{Z'}$ does not generate enough violation of the CKM unitarity according to the experimental result in Eq. (\ref{ckm}). Thus, this region is also not favored.
The lower bound on $m_Q$ set by the current constraints on the $b\rightarrow s \gamma$ decay (the light green region) 
and the CKM unitarity is 860 GeV.
This bound is expected to rise up to nearly 
2130 GeV after the Belle II experiment imposes a more restrictive constraint (\ref{bsg_belle2}) (the dark green region).
Taking into account the constraints (\ref{BRK})-(\ref{BRKse}) from the data of the semileptonic decays $b\rightarrow s \ell^+ \ell^-$ (the yellow band),
we obtain even more severe allowed range for $m_{Z'}$ and $m_Q$.
The current allowed ranges for $m_{Z'}$ and $m_Q$ are
[150, 1000] GeV and
[2000, 2615] GeV, respectively.
We observe that this viable region already satisfies the current LHC lower limit (\ref{mQ_limit}) on the vectorlike quark mass (the horizontal dash-dotted line).
The allowed ranges for these two parameters after the Belle II experiment are expected to be reduced to
[152, 816] GeV and
[2124, 2615] GeV.

%%%%%%%%%%%%%%%%%%%%%%%%%%%%%%%%%%
\begin{figure}[h!]
\begin{center}
\includegraphics[scale=0.65]{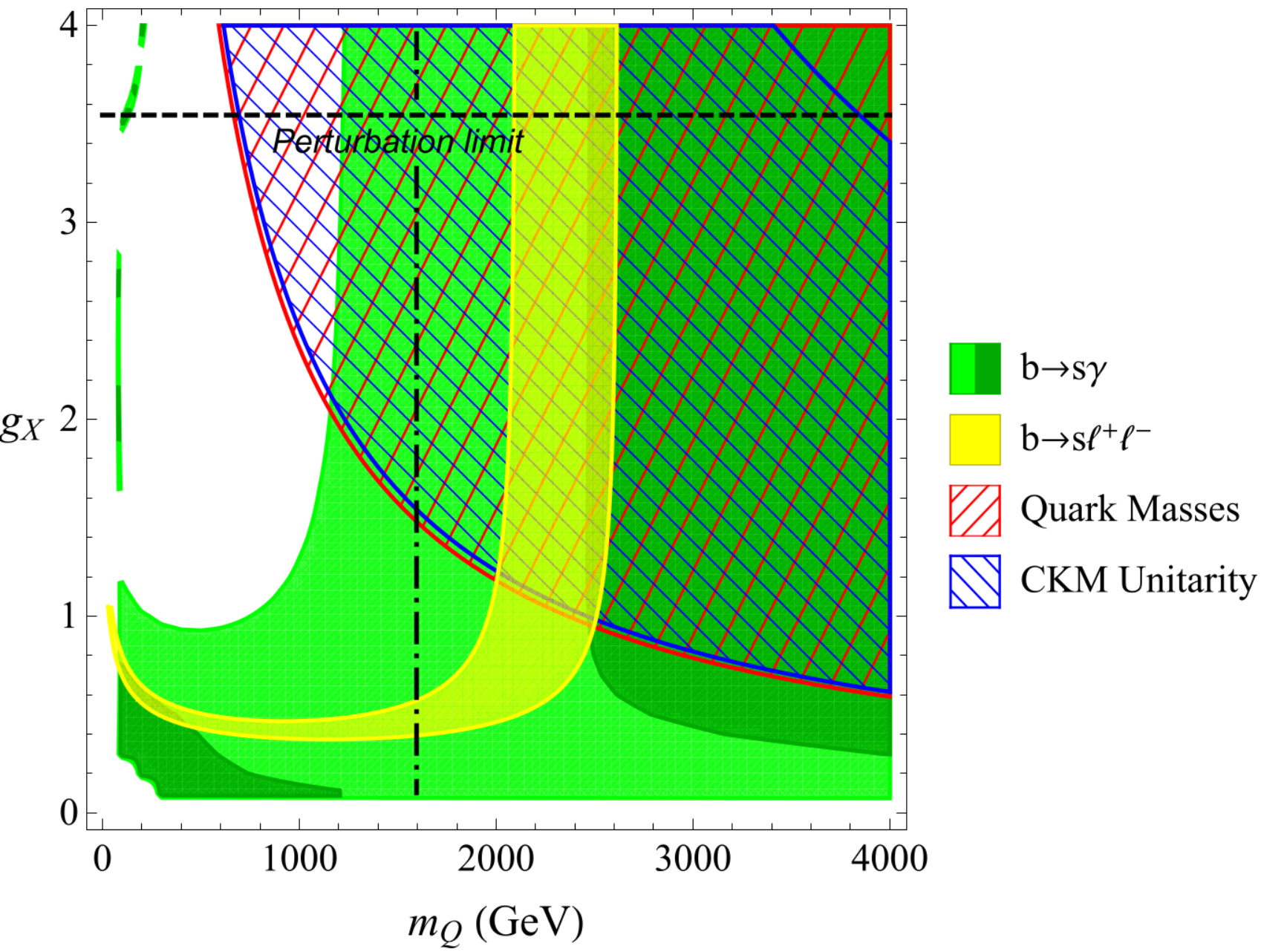}
\caption{Constraints on the $(m_Q,g_X)$ plane
 when
$m_{Z'}=700$ GeV,
%$g_X = 2$, 
%$\lambda_\phi = 3$,
%$w_1 = 0$,
$w_2 = 2.45$, 
$w_3 = 0.25$.
The color codes are the same as those in Figure \ref{mzpmq}.
The horizontal black dashed line indicates the perturbation limit (\ref{perturbation}).
The vertical dash-dotted line corresponds to the lower bound (\ref{mQ_limit}) on $m_Q$.
}
\label{mqgx}
\end{center}
\end{figure}
%%%%%%%%%%%%%%%%%%%%%%%%%%%%%%%%%%

In Figure \ref{mqgx}, we show the 
2$\sigma$ constraints on the plane of ($m_Q, g_X$) with fixed values of other parameters:
$m_{Z'} = 700$ GeV,
%$\lambda_\phi = 3$,
%$w_1 = 0$,
$w_2 = 2.45$, and
$w_3 = 0.25$.
Using the same color codes as those in Figure \ref{mzpmq},
the red hatched region satisfies the constraints on the SM quark masses, while the blue back-hatched region indicates the constraint (\ref{ckm}) from the measurements of the first-row elements of the CKM.
We observe that these two constraints are compatible, with the one on the CKM unitarity violation being more severe.
It excludes a significant part of the parameter space with small values of $m_Q$ and $g_X$ that corresponds to the region with too large mixing between the SM and the vectorlike quarks.
The region with too small mixing, corresponding to too large $g_X$ and $m_Q$ (the upper right corner of the plot), is also excluded since it could not generate enough unitarity violation according to Eq. (\ref{ckm}).
The constraints on the $b\rightarrow s \gamma$ decay derived from the current experimental data (\ref{bsg_now}) and the expected Belle II data
(\ref{bsg_belle2}) are depicted in this plot by the light and dark green regions, respectively.
While the current bounds (\ref{bsg_now}) exclude the white region on the left with both 
$m_Q \lesssim 1100$ GeV and 
$g_X \gtrsim \mathcal{O}(1)$ simultaneously,
the expected result from Belle II experiment will be able to set a more stringent lower limit on $m_Q$ to be about 
2460 GeV.
The yellow 
band in this figure indicates the combined constraint from the
$b \rightarrow s \ell^+ \ell^-$ transitions ((\ref{BRK})-(\ref{BRKse})).
The overlap between this 
band and the constraint on the CKM unitarity violation severely restrict the allowed range for $m_Q$, which is  
[2090, 2610] GeV for this benchmark point.
As shown in the figure, this range obviously fulfills the current LHC requirement (\ref{mQ_limit}) on the vectorlike quark mass (the vertical dash-dotted line).
Since the dark green area only marginally overlaps the yellow band, the Belle II experiment will be able to test the allowed region in the near future.

%%%%%%%%%%%%%%%%%%%%%%%%%%%%%%%%%%
\begin{figure}[h!]
\begin{center}
\includegraphics[scale=0.65]{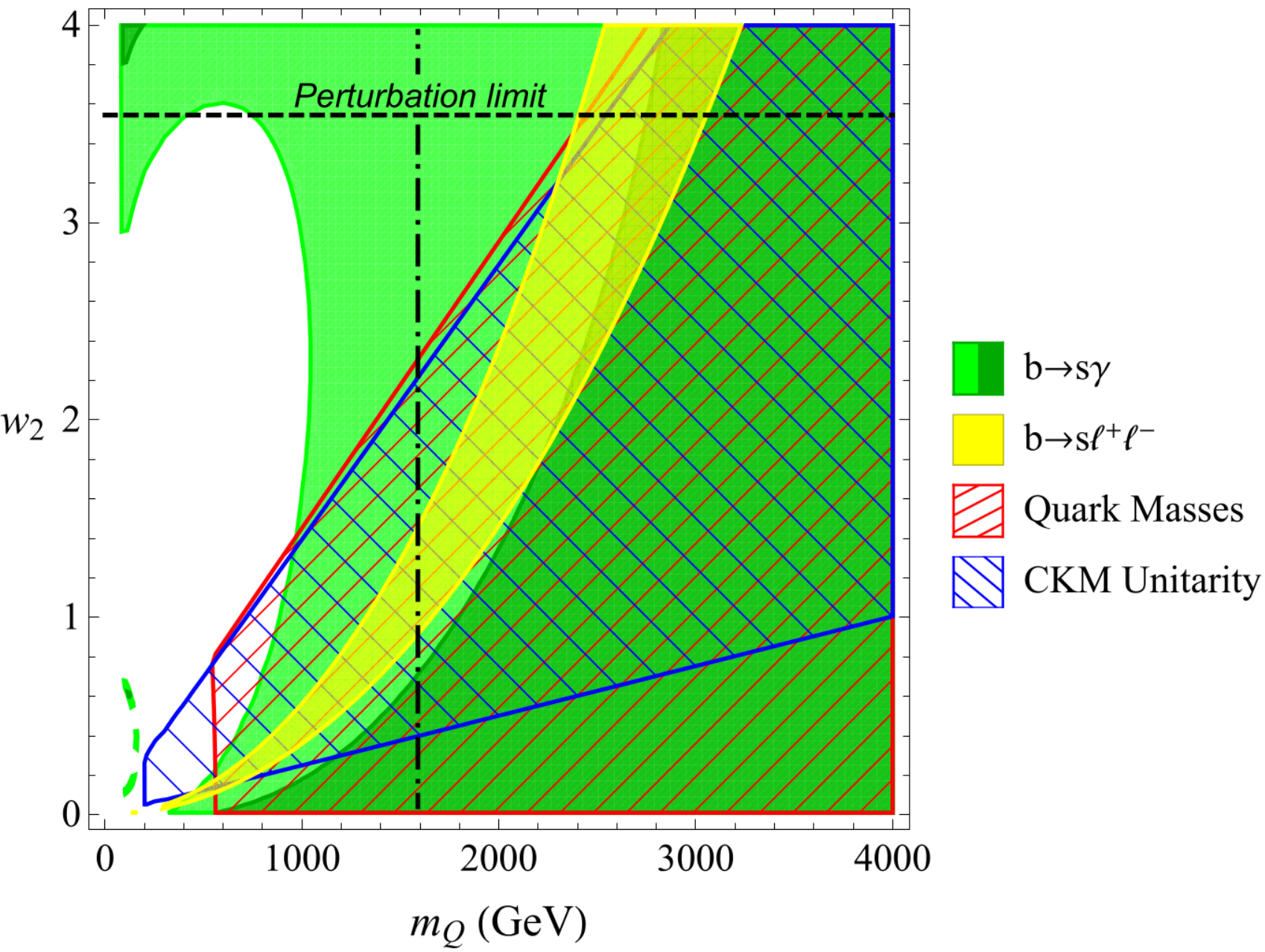}
\caption{Constraints on the $(m_Q,w_2)$ plane
 when
$m_{Z'}=700$ GeV,
$g_X = 1.4$, 
%$\lambda_\phi = 3$,
%$w_1 = 0$,
%$w_2 = 1.55$, 
$w_3 = 0.25$.
The color codes are the same as those in Figure \ref{mzpmq}.
The horizontal black dashed line indicates the perturbation limit (\ref{perturbation}).
The vertical dash-dotted line corresponds to the lower bound (\ref{mQ_limit}) on $m_Q$.}
\label{mqw2}
\end{center}
\end{figure}
%%%%%%%%%%%%%%%%%%%%%%%%%%%%%%%%%%

The constraints on the ($m_Q, w_2$) plane
are shown in Figure \ref{mqw2}.
Here, the benchmarks for other parameters are chosen to be 
$m_{Z'}=700$ GeV,
$g_X = 1.4$, 
%$\lambda_\phi = 3$,
%$w_1 = 0$,
%$w_2 = 1.55$, 
$w_3 = 0.25$.
The region with small $w_2$ corresponding to very little mixing between the SM and vectorlike quarks 
does not satisfy the constraint (\ref{ckm}) on the CKM unitarity violation.
Therefore, this constraint represented by the blue back-hatched region requires $w_2$ to have sizable values,
when $m_Q$ is fixed.
Due to the interplay between $w_2$ and $m_Q$ in terms of their effects on the 
BR$(b \rightarrow s \gamma)$ and 
the quark masses,
a large value of $w_2$ is not enough if the vectorlike quarks are relatively light.
It is observed that the combination of these two constraints 
(the red hatched and the green areas)
implies that $m_Q$ must be larger than about 560 GeV.
The allowed regions on the ($m_Q,w_2$) plane are further restricted
when we take into account the constraints (\ref{BRK})-(\ref{BRKse})
on the semileptonic rare decays of $B$ mesons (the yellow region).
The combination of the CKM unitarity violation constraint (\ref{ckm}) and
this set of constraints imposes a lower limit of about 0.16
on the parameter $w_2$.
Taking into account LHC lower bound (\ref{mQ_limit}) on the vectorlike quark mass (the vertical dash-dotted line),
the lower limit on $w_2$ is improved to be  about 0.9.
In the near future, the Belle II experiment is expected to push the lower bound of $m_Q$ up to about 
2090 GeV, and the lower bound of $w_2$ to 
about 1.6
for this benchmark when superimposing the dark green area and the yellow one.
The upper bound on $m_Q$, in this case, is about 3060 GeV that is determined by the perturbation limit (\ref{perturbation}) on the Yukawa coupling $w_2$ (the horizontal black dashed line).

%%%%%%%%%%%%%%%%%%%%%%%%%%%%%%%%%%
\begin{figure}[h!]
\begin{center}
\includegraphics[scale=0.65]{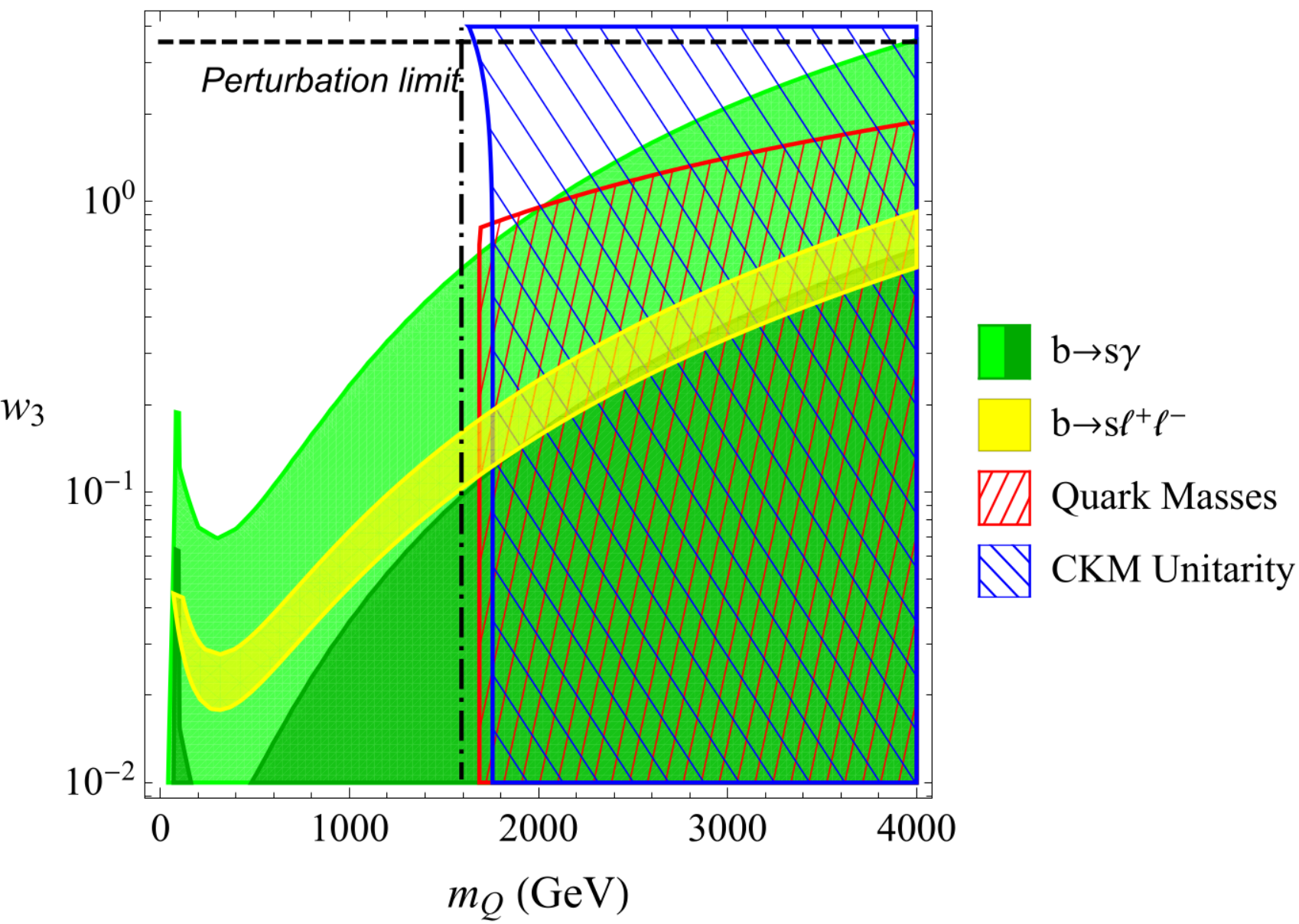}
\caption{Constraints on the $(m_Q,w_3)$ plane
 when
$m_{Z'}=700$ GeV,
$g_X = 1.4$, 
%$\lambda_\phi = 3$,
%$w_1 = 0$,
$w_2 = 2.45$. 
%$w_3 = 0.145$.
The color codes are the same as those in Figure \ref{mzpmq}.
The vertical dash-dotted line corresponds to the lower bound (\ref{mQ_limit}) on $m_Q$.}
\label{mqw3positive}
\end{center}
\end{figure}
%%%%%%%%%%%%%%%%%%%%%%%%%%%%%%%%%%

In Figure \ref{mqw3positive}, we show the allowed regions with respect to the considered constraints on the plane ($m_Q, w_3$) for the case with
$m_{Z'}=700$ GeV,
$g_X = 1.4$, and
%$\lambda_\phi = 3$,
%$w_1 = 0$, and
$w_2 = 2.45$.
In this case, the lower bound for $m_Q$ is determined by the constraint on the CKM unitarity violation (\ref{ckm}) (the blue back-hatched region) to be about 1760 
GeV for 
$w_3 \lesssim 0.9$.
It is due to the fact that smaller $m_Q$ will lead to too much mixing between the SM and the vectorlike quark.
For 
$w_3 \gtrsim 0.9$, the constraint on the SM quark masses becomes severe and rules out most of the parameter space because of the same reason as the above case with small $m_Q$.
This constraint is even more severe than
the current constraint (\ref{bsg_now}) on the $b \rightarrow s \gamma$ decay (the light green region).
In this figure, we see that once the constraints on the CKM unitarity and the SM quark masses are imposed, the allowed parameter region also satisfies the LHC lower limit (\ref{mQ_limit}) on $m_Q$ (the vertical dash-dotted line).
The constraints (\ref{BRK})-(\ref{BRKse})
on the semileptonic decay  $b \rightarrow s \ell^+ \ell^-$ play an important role in excluding the parameter space such that only a thin yellow band
survives on this plot.
Although this yellow band
satisfies the current constraint on 
BR$(b\rightarrow s \gamma)$, the expected result at the Belle II experiment (the dark green region) will be able to rule out a significant part of this band.

%%%%%%%%%%%%%%%%%%%%%%%%%%%%%%%%%%
\begin{figure}[h!]
\begin{center}
\includegraphics[scale=0.65]{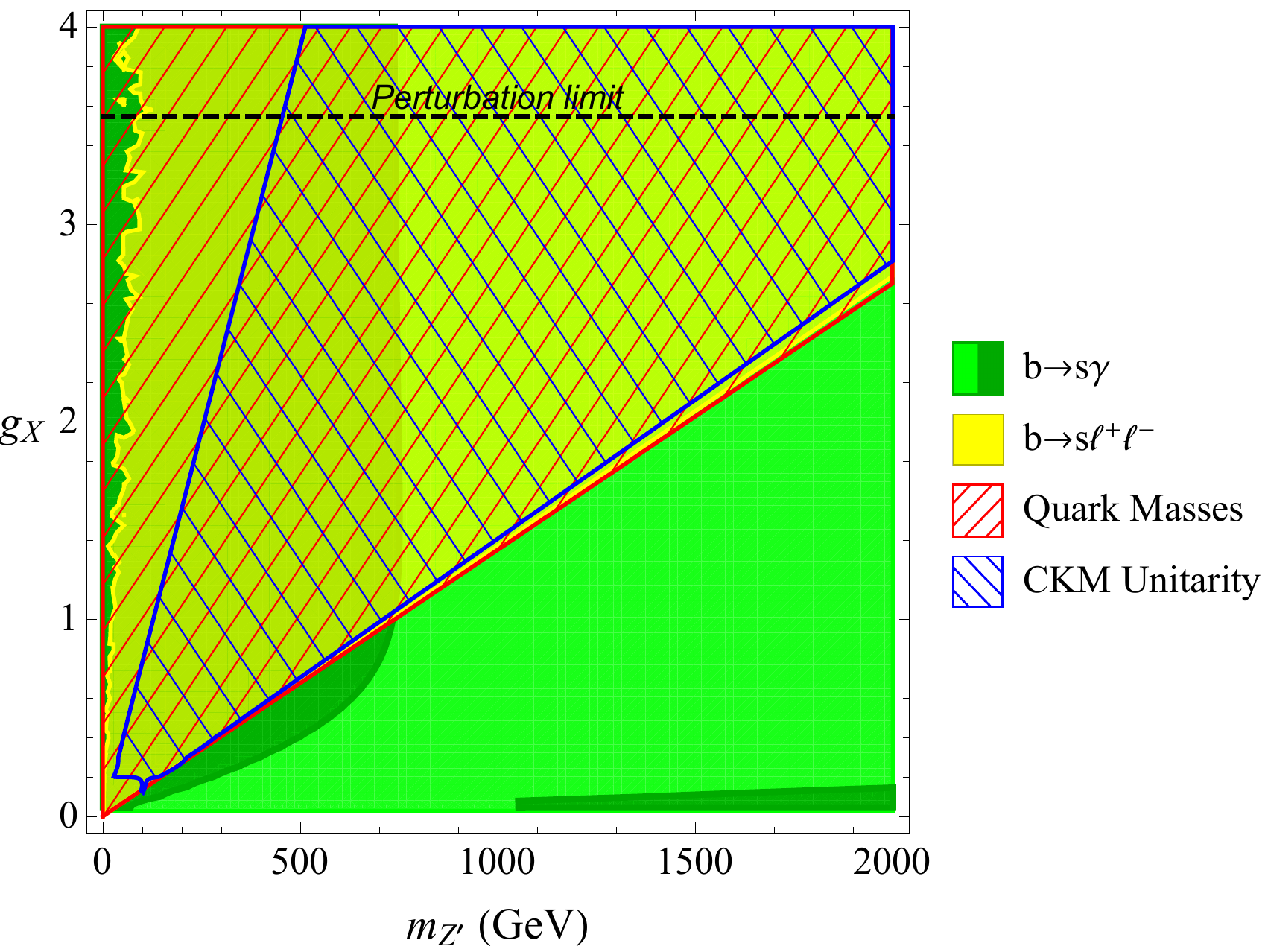}
\caption{Constraints on the $(m_{Z'},g_X)$ plane
 when
$m_{Q}=2500$ GeV,
%$g_X = 2$, 
%$\lambda_\phi = 3$,
%$w_1 = 0$,
$w_2 = 2.45$, 
$w_3 = 0.25$.
The color codes are the same as those in Figure \ref{mzpmq}.
The horizontal black dashed line indicates the perturbation limit (\ref{perturbation}).
}
\label{mzpgx}
\end{center}
\end{figure}
%%%%%%%%%%%%%%%%%%%%%%%%%%%%%%%%%%

The allowed regions on the plane ($m_{Z'}, g_X$) are plotted in Figure \ref{mzpgx}
in the case with 
$m_{Q}=2500$ GeV,
%$g_X = 2$, 
%$\lambda_\phi = 3$,
%$w_1 = 0$,
$w_2 = 2.45$, and
$w_3 = 0.25$.
With this choice of benchmark, the considered parameter space in the figure satisfies the current constraint
(\ref{bsg_now}) on the $b\rightarrow s \gamma$ decay.
However, the upcoming result (\ref{bsg_belle2}) at the Belle II experiment (the dark green region)
will be expected to set a severe upper bound on $m_{Z'}$ to be about 750 GeV.
On the one hand, the constraint (\ref{ckm}) on the CKM unitarity violation (the blue back-hatched region) imposes a lower bound of about 
128 GeV
and the upper bound of about 714 GeV
on the ratio 
$\frac{m_{Z'}}{g_X}$.
The regions with the ratio 
outside this range are excluded because 
they generate either too small mixing or too large mixing
between the SM and the vectorlike quarks.
For this benchmark,
the constraint on the SM quark masses (the red hatched region) and the combination of constraints (\ref{BRK})-(\ref{BRKse})
on the semileptonic decays $b \rightarrow s \ell^+ \ell^-$ (the yellow region) 
are also fulfilled once the ratio
$\frac{m_{Z'}}{g_X}$ stays in the above range.

%%%%%%%%%%%%%%%%%%%%%%%%%%%%%%%%%%
\begin{figure}[h!]
\begin{center}
\includegraphics[scale=0.65]{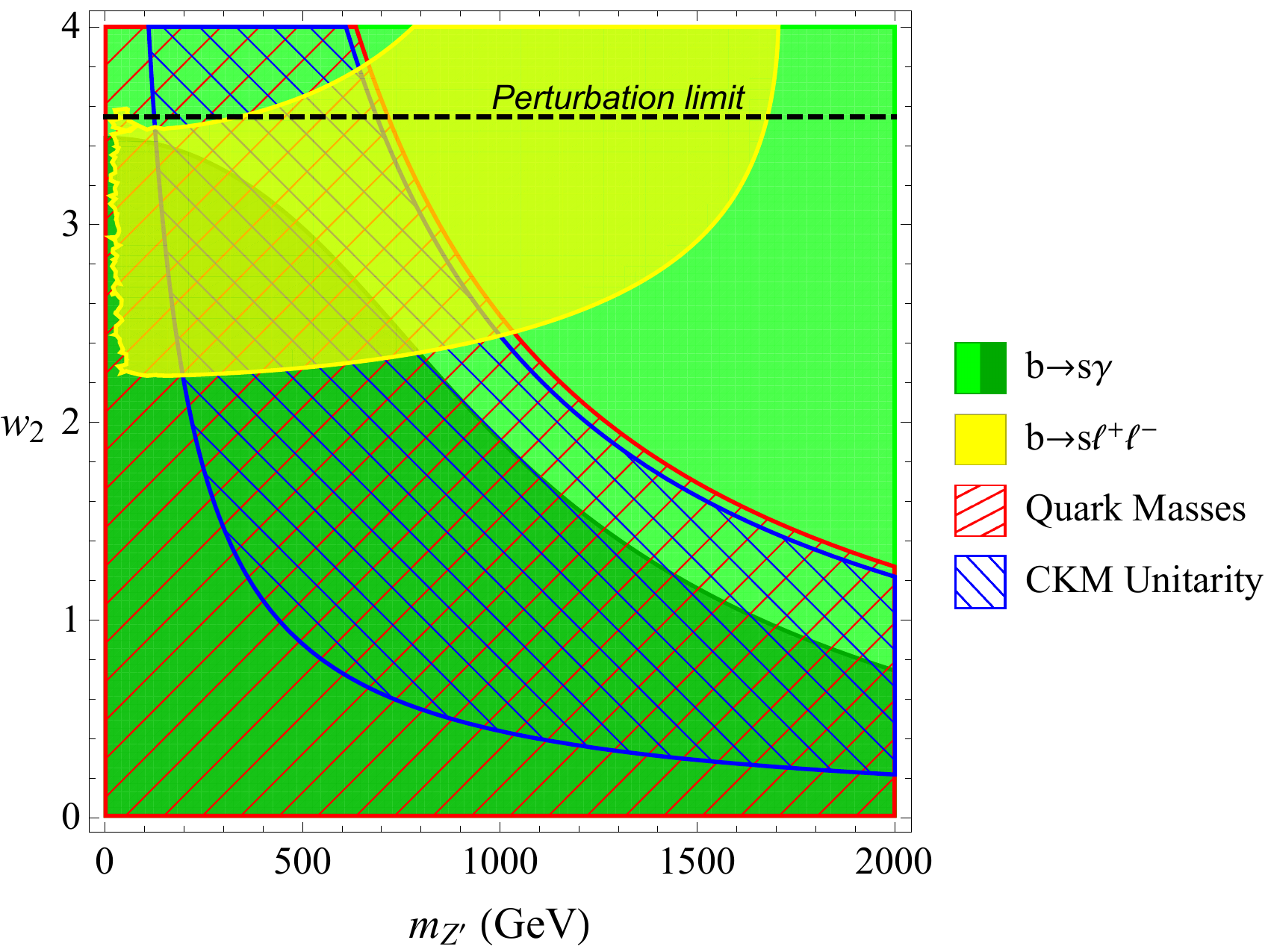}
\caption{Constraints on the $(m_{Z'},w_2)$ plane
 when
$m_{Q}=1700$ GeV,
$g_X = 1.4$, 
%$\lambda_\phi = 3$,
%$w_1 = 0$,
%$w_2 = 1.55$, 
$w_3 = 0.25$. 
The color codes are the same as those in Figure \ref{mzpmq}.
}
\label{mzpw2}
\end{center}
\end{figure}
%%%%%%%%%%%%%%%%%%%%%%%%%%%%%%%%%%

In Figure \ref{mzpw2},
the constraints on the plane ($m_{Z'},w_2$) are shown in the case with
$m_{Q}=2500$ GeV,
$g_X = 1.4$, 
%$\lambda_\phi = 3$,
%$w_1 = 0$, and
%$w_2 = 1.55$, 
$w_3 = 0.25$.
In this plot, we can see that for larger $m_{Z'}$, the constraint on the CKM unitarity violation prefers smaller values of $w_2$, while the constraints on the semileptonic decays of $B$ meson prefer larger values of $w_2$.
The points in the region with small $w_2$ and $m_{Z'}$ correspond to too small mixing between the SM and the vectorlike quarks.
Therefore, they can not explain the CKM unitarity violation in the first row.
This results in the excluded region below the blue back-hatched area.
Although the current $b\rightarrow s\gamma$ constraint (\ref{bsg_now}) is not severe for this benchmark, the Belle II experiment is expected to make a significant contribution in excluding a large portion of the parameter space.
As we can see in this figure, the dark green region corresponding to Eq. (\ref{bsg_belle2}) becomes 
smaller than the red hatched region allowed by the quark mass constraint.
The presently allowed regions on this plane are set by the constraints
(\ref{BRK})-(\ref{BRKse}) on the $b\rightarrow s \ell^+ \ell^-$ transitions (the yellow region), the constraint (\ref{ckm}) from the CKM unitarity violation (the blue back-hatched region), and the perturbation limit on $w_2$.
With the given choice of other parameters as above, the Yukawa coupling $w_2$ is restricted to stay within a narrow range [2.24, $\sqrt{4\pi}$],
while the allowed range for $m_{Z'}$ is [125, 1000] GeV.
After the Belle II experiment finishes, this allowed region will be reduced by a factor of about one half.

%%%%%%%%%%%%%%%%%%%%%%%%%%%%%%%%%%
\begin{figure}[h!]
\begin{center}
\includegraphics[scale=0.65]{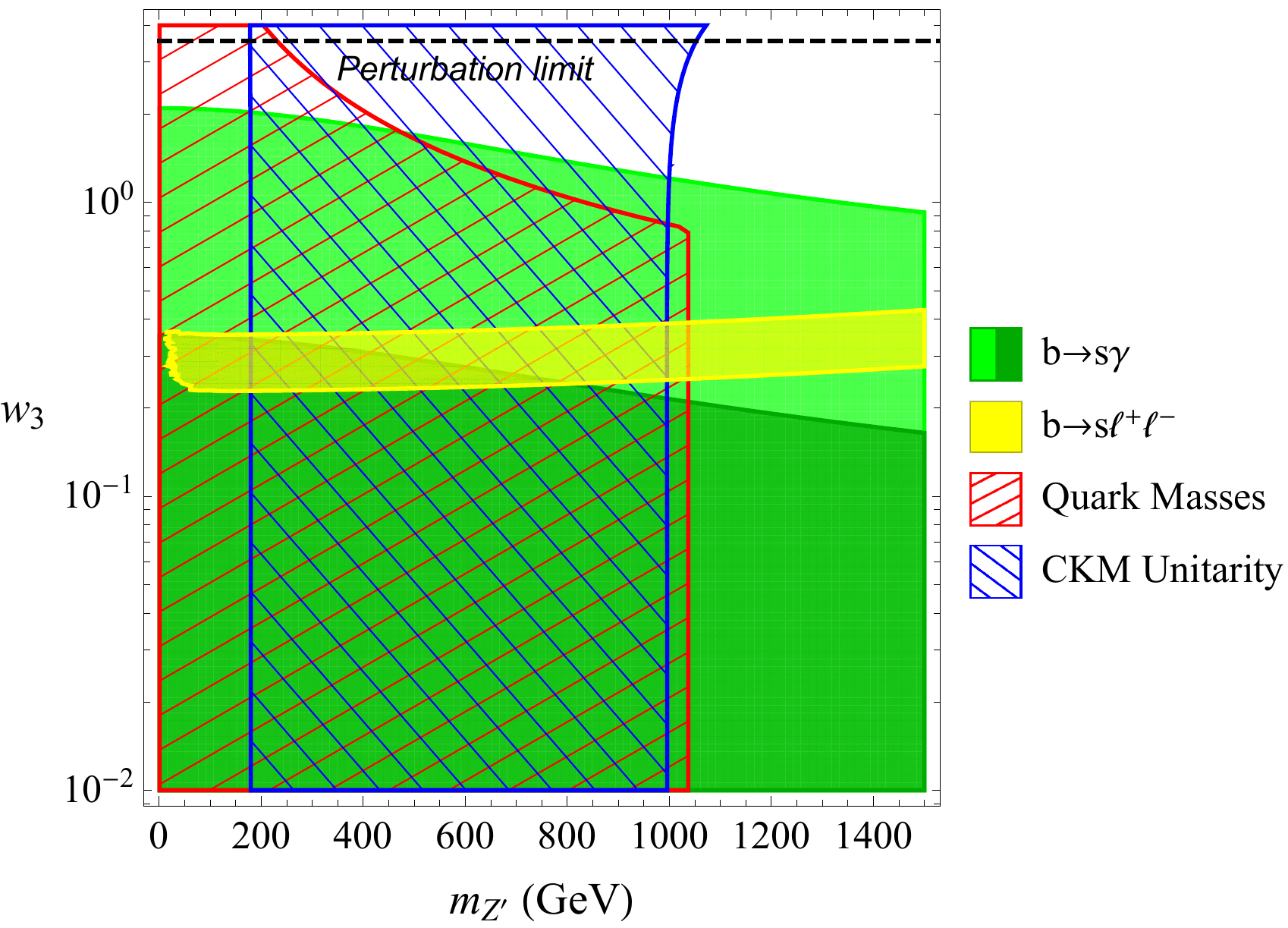}
\caption{Constraints on the $(m_{Z'},w_3)$ plane
 when
$m_{Q}=2500$ GeV,
$g_X = 1.4$, 
%$\lambda_\phi = 3$,
%$w_1 = 0$,
$w_2 = 2.45$.  
%$w_3 = 0.145$.
The color codes are the same as those in Figure \ref{mzpmq}.
}
\label{mzpw3}
\end{center}
\end{figure}
%%%%%%%%%%%%%%%%%%%%%%%%%%%%%%%%%%

In Figure \ref{mzpw3}, the considered constraints are depicted on the plane ($m_{Z'}, w_3$).
The free inputs are chosen to be
$m_{Q}=2500$ GeV,
$g_X = 1.4$, and 
%$\lambda_\phi = 3$,
%$w_1 = 0$, and
$w_2 = 2.45$.
For this benchmark, the CKM unitarity violation constraint (the blue back-hatched region) requires the $Z'$-boson mass to be within the range
[180, 1000] GeV.
The constraints on the semileptonic decays $b\rightarrow s \ell^+ \ell^-$ (the yellow strip) restrict the values of $w_3$ to be about $0.22 - 0.40$
in this case.
The expected result at the Belle II experiment will provide us with a stringent constraint on the $b\rightarrow s \gamma$ decay (the dark green region).
According to that the allowed range of $m_{Z'}$ will be reduced to 
[180, 820] GeV
 for this benchmark point.

%%%%%%%%%%%%%%%%%%%%%%%%%%%%%%
\section{Conclusion}
%%%%%%%%%%%%%%%%%%%%%%%%%%%%%%

In the considered model with a new sector consisting of vectorlike fermions and two scalars charged under an extra $U(1)_X$ symmetry, the exotic Yukawa interactions between this sector and the SM fermions are essential to address various experimental anomalies such as 
the muon $g-2$ 
and the semileptonic decays of $B$ mesons.
Within this context, we have analytically calculated 
the new physics contributions to the Wilson coefficient $C_7$ in the effective Hamiltonian.
Based on that, the dependence of the $b\rightarrow s \gamma$ decay rate on the free parameters is obtained.
We have shown that this model is possible to explain the measured CKM unitarity violation and the current data relevant to the $b \rightarrow s \gamma$ transition at the same time, 
while predicting 
other flavor observables relating to the $b \rightarrow s \ell^+ \ell^-$ processes and the SM quark masses compatible with their updated measurements.
In our analysis, the muon anomalous magnetic moment is ensured to be
 consistent with the recent data measured by the Muon $g-2$ experiment.
By investigating the space of input parameters, the allowed region satisfying various constraints has been identified.
Together with the combined constraint on the semileptonic decays of $B$ mesons, the constraint on the violation of the CKM unitarity plays an important role in pinpointing the viable parameter space.
Taking into account the recent LHC searches for the vectorlike quarks,
the impact of the expected outcome at the Belle II experiment on the $b\rightarrow s \gamma$ decay has been analyzed in detail.
The result has showed that this foreseen constraint will be able to exclude a significant portion of the currently allowed parameter regions in the near future.

%\end{fmffile}

%%%%%%%%%%%%%%%%%%%%%%%%%%%

%%%%%%%%%%%%%%%%%%%%%%%%%%%%%%
\section*{Acknowledgment}
%%%%%%%%%%%%%%%%%%%%%%%%%%%%%%

Sang Quang Dinh was funded by Vingroup JSC and supported by the Master, PhD Scholarship Programme of Vingroup Innovation Foundation (VINIF), Institute of Big Data, code VINIF.2021.TS.037.

%%%%%%%%%%%%%%%%%%%%%%%%%%%%%%
\section*{Appendix}
%%%%%%%%%%%%%%%%%%%%%%%%%%%%%%

\appendix

\section{New physics contributions to $C^{(')}_{9,10}$}
\label{C9-10}

The new physics contributions to the Wilson coefficients $C_{9,10}^{(')}$ in the absence of the gauge kinetic mixing are given as \cite{Dinh:2020inx}
\begin{eqnarray}
C^{\text{NP}}_9	&=&	
	\frac{g_X^2 A}{ m_{Z'}^2}
	\Lambda_\text{SM}^2
	\frac{|V_{tb} V_{ts}^*|}{V_{tb} V_{ts}^*} A_{bs}	,	\\
C'^{\text{NP}}_9	&=&	
	\left[
		\frac{g_X^2 A}{m_{Z'}^2}  	
		-
		\frac{g_2^2 \left( 
		-\frac{1}{2} + \frac{1}{3} \sin^2 \theta_W
		\right) A^Z}{2 m_Z^2 \cos \theta_W }
	\right] 
	\Lambda_\text{SM}^2
	\frac{|V_{tb} V_{ts}^*|}{V_{tb} V_{ts}^*} B_{bs}	,	 
	\\
C^{\text{NP}}_{10}	&=&	
	\frac{g_X^2 B}{ m_{Z'}^2} 
	\Lambda_\text{SM}^2
	\frac{|V_{tb} V_{ts}^*|}{V_{tb} V_{ts}^*} A_{bs}	,\\
C'^{\text{NP}}_{10}	&=&
	\left[
		\frac{g_X^2 }{m_{Z'}^2} B 
	-
		\frac{g_2^2 \left( 
		-\frac{1}{2} + \frac{1}{3} \sin^2 \theta_W
		\right) B^Z}{2 m_Z^2 \cos \theta_W }
	\right] 
	\Lambda_\text{SM}^2
	\frac{|V_{tb} V_{ts}^*|}{V_{tb} V_{ts}^*} B_{bs}	,
\end{eqnarray}
where the intermediate notations $A_{bs}$, $B_{bs}$, $A$, $A^Z$, $B$, and $B^Z$ are defined as follows
\begin{eqnarray}
A_{bs}	&=&	
	\left( V^d_L \right)_{34}
	\left( V_L^{d} \right)_{24}^* \, ,	\\
B_{bs}	&=&
	\left( V^d_R \right)_{34}
	\left( V_R^{d} \right)_{24}^* \,	.
\end{eqnarray}
\begin{eqnarray}
A	(q^2)	&=&	
	\frac{|y_\ell|^2 }{32\pi^2}	 
	 (f_A + g_A)	 ,	\\
%\end{eqnarray}
%
%\begin{eqnarray}
A^Z (q^2)	&=&		
	\frac{1}{\cos \theta_W} 
	\left( -\frac{1}{4} + \sin^2 \theta_W \right)	 
	+ \, \frac{|y_\ell|^2}{32\pi^2}
	\frac{g_A}{\cos \theta_W}
	\left( -\frac{1}{2} + \sin^2 \theta_W \right)  ,		\\
%\end{eqnarray}
%
%\begin{eqnarray}
B (q^2)	&=&	
	\frac{|y_\ell|^2}{32\pi^2}
	( - f_B +	g_B )  ,	\\
%\end{eqnarray}
%
%\begin{eqnarray}		
B^Z (q^2)	&=&	
	\frac{1}{4\cos \theta_W}  	
	+ \frac{|y_\ell|^2}{32\pi^2}
	\frac{	g_B}{\cos \theta_W}
	\left( -\frac{1}{2} + \sin^2 \theta_W \right) .
\end{eqnarray}
In these above formulas, the loop functions $f_A$, $g_A$, $f_B$, and $g_B$ are given by
\begin{eqnarray}
f_A	&=&
	\int dx dy dz \delta(1-x-y-z)	
	\left\{
	\frac{\ln \left[ (\tau z +x+y+ \delta x) (\tau z +x+y+ \delta y) \right] }{2}
	\right. 	\nonumber	\\
&&	 
	\qquad \qquad \qquad	
	\left.
	- \; \frac{m_\ell^2}{m_{\chi_r}^2}
	z(1-z)
	\left[ 
		\frac{1}{\tau z +x+y+ \delta x} + 
		\frac{1}{\tau z +x+y+ \delta y}
	\right]
	\right\}	\, ,	\\
%\end{eqnarray}
%
%\begin{eqnarray}
g_A	&=&
	\int dx dy dz \delta(1-x-y-z)
	\left\{
	-\frac{ 
	\ln \left[ (\tau (x+y) +z)  (\tau(x+y) +z + \delta z) \right] }{2}
	\right.	\nonumber	\\
&&	
	\qquad \qquad
	\left.
	+ \, \frac{z^2 m_\ell^2 + xyq^2 + m_L^2}{2m_{\chi_r}^2}
	\left[
		\frac{1}{\tau (x+y) + z} +
		\frac{1}{\tau (x+y) + z + \delta z }
	\right]
	\right\}	\, ,	\\
%\end{eqnarray}
%
%\begin{eqnarray}
f_B	&=&	
	\int dx dy dz \delta(1-x-y-z)
	\frac{\ln \left[ (\tau z +x+y+ \delta x) (\tau z +x+y+ \delta y) \right]}{2}	\, ,	\\
%\end{eqnarray}
%
%\begin{eqnarray}
g_B	&=&	
	\int dx dy dz \delta(1-x-y-z) 
	\left\{
	\frac{\ln \left[ (\tau (x+y) +z) (\tau (x+y) +z+ \delta z) \right] }{2}
	\right.	\nonumber	\\
&&	
	\left.
	\qquad \qquad 
	+ \, \frac{z^2 m_\ell^2 - xyq^2 - m_L^2}{2 m_{\chi_r}^2}
	\left[
		\frac{1}{\tau (x+y) +z} +
		\frac{1}{\tau (x+y) +z+ \delta z}
	\right]
	\right\}	\, ,
\end{eqnarray}
as the results of the Feynman parameterization.
In these formulas, $y_\ell$ is the exotic Yukawa coupling of one of the SM charge leptons $\{e, \mu, \tau\}$.

\section{New physics contributions to $g_\ell - 2$}
\label{appendix_g-2}

The new physics contributions to the anomalous magnetic moment of charged leptons are given as 
\cite{Dinh:2020inx, Belanger:2015nma}

\begin{eqnarray}
\Delta a_\ell^\text{NP}	&=&
	\frac{|y_\ell|^2 m_\ell^2}{32\pi^2 m_{\chi_r}^2}
	\left[
		F_g(\tau) + 
		\left( \frac{1}{1+\delta}	\right)
		F_g \left( \frac{\tau}{1+\delta} \right)
	\right] 
\label{amu_k}
\end{eqnarray}
where
\begin{eqnarray}
\tau	&=&	
	\frac{m_L^2}{m_{\chi_r}^2} ,	\\
\delta	&=&
	\frac{m_{\chi_i}^2}{m_{\chi_r}^2} - 1	.	
	\label{delta_define}	
\end{eqnarray}
Here, the loop function $F_g(x)$ is defined as
\begin{eqnarray}
F_g (x)	&=&
	\frac{1}{6(1-x)^4}
	\left(
		6x \ln x + x^3 -6x^2 + 3 x +2
	\right)	.
\end{eqnarray}

%%%%%%%%%%%%%%%%%%%%%%%%%%%%%%%%%

\end{document}